\documentclass[a4paper,fleqn,usenatbib]{mnras}
\pdfoutput=1

\usepackage[T1]{fontenc}
\usepackage{ae,aecompl}

\usepackage{pdflscape} 

\usepackage{graphicx}	
\usepackage{amsmath}	
\usepackage{amssymb}	

\usepackage{mathrsfs}
\usepackage[english]{babel}

\usepackage[usenames]{color}

\newcommand{\f}[2]{\frac{#1}{#2}} 

\newcommand{\athena}{\textsc{Athena{\scriptsize ++ }}}
\newcommand{\athenas}{\textsc{Athena{\scriptsize ++}}'s }

\usepackage[usenames]{color}

\newcommand{\beq}{\begin{equation}}
\newcommand{\seq}{\end{equation}}
\let\f=\ff
\newcommand{\f}[2]{\frac{#1}{#2}} 

\newcommand{\gv}[1]{\ensuremath{\mbox{\boldmath$ #1 $}}} 

\title[]
{Magnetothermal disk winds in X-ray binaries:\\ poloidal magnetic fields suppress thermal winds}

\author[Waters \&Proga]{
Tim Waters$^1$\thanks{E-mail: waters@lanl.gov} and 
Daniel Proga$^2$
\\
$^1$Theoretical Division, Los Alamos National Laboratory, Los Alamos, NM, 87545, USA\\
$^2$Department of Physics \& Astronomy, University of Nevada, Las Vegas, 
  4505 S. Maryland Pkwy, Las Vegas, NV, 89154-4002, USA\\
}

\date{Accepted XXX. Received YYY; in original form ZZZ}

\pubyear{2016}

\begin{document}
\label{firstpage}
\pagerange{\pageref{firstpage}--\pageref{lastpage}}
\maketitle

\begin{abstract}
Magnetic, radiation pressure, and thermal driving are the three mechanisms capable of launching accretion disk winds.   
In X-ray binaries, radiation pressure is often not significant, as in many systems the luminosity is too low for driving due to continuum transitions yet too high for driving due to line transitions.  This leaves thermal and magnetic driving as the contender launching mechanisms in these systems.  Using \athena, we perform axisymmetric ideal MHD simulations that include radiative heating and cooling processes appropriate for Compton heated winds to show that the inclusion of magnetic fields into a thermally driven wind has the opposite effect of what one might expect: rather than provide a velocity boost, the thermal wind is suppressed in low plasma beta regions where the field lines are strong enough to reshape the direction of the flow.  Our analysis reveals that magneto-centrifugal launching is present but weak, while the reduction in wind velocity is not due to the change in gravitational potential through the magnetically imposed streamline geometry, but rather due to the increased flow tube area just above the surface of the disk, which is less conducive to acceleration.  Our results suggest that for magnetothermal wind models to be successful at producing fast dense outflows in low mass X-ray binaries, the winds must be magnetically launched well within the Compton radius.

\end{abstract}


\section{Introduction}
Accretion disk winds, like stellar winds, can be thermally, radiatively, or magnetically launched.  
Thermal launching in stellar winds ultimately owes its source of heat to magnetic effects 
(e.g., dissipation due to magnetic reconnection), whereas in disk winds the 
source of heat can be the radiation field (e.g., X-ray irradiation from the inner regions of the disk).  
Regardless of the source of heating, the wind is deemed thermally driven when the gas pressure force is responsible for launching the outflow.  Radiative driving involves a direct transfer of momentum from photons to the gas.  It can be effective either in the form of continuum driving when the luminosity is high, in which case the momentum transfer is from electron scattering, bound-free absorption, or dust absorption; or via `line driving', which is momentum transfer from photons scattering off of many individual spectral lines.   

Much insight into the physics of magnetic driving resulted from the model of Blandford \& Payne (1982).  The theory of magneto-centrifugal driving was first developed by Weber \& Davis (1967) and applied to stellar winds, where it is known as magnetic rotator theory.  This launching mechanism requires that the magnetic field apply a sufficient torque to the gas.  In the context of disks,
provided the poloidal field is not too vertical --- it must be inclined less than $60^\circ$ to the surface of the disk, in the case of negligible pressure (Blandford \& Payne 1982) --- the centrifugal force supplied by the rotation of the disk can be sufficient to overcome gravity and the gas can be accelerated outward.  
 
In stellar winds, magnetic driving is strong only near the equatorial plane of the star, and so this theory is typically invoked along with a `primary wind mechanism', the primary wind being thermally or radiatively launched (e.g., Belcher \& MacGregor 1976; Nerney 1980; Poe et al. 1989).  The basic findings have been that magnetic forces help to radially accelerate the flow, advect angular momentum outward, and to increase the mass loss rate (e.g., Cassinelli 1990).  Magneto-centrifugal launching actually corresponds to an extreme case of magnetic rotator theory (Michel 1969; Belcher \& MacGregor 1976), and on this basis we are left with the expectation that magnetic driving will not impede a (thermally or radiatively driven) `primary' disk wind.  

In actual accretion disk environments, more than one wind driving mechanism is likely operating.  X-ray binaries, for example, which are tight binary systems featuring mass transfer from a secondary star onto a neutron star or stellar mass black hole, can have high X-ray luminosities and are host to radio jets, and thus in principle all three launching mechanisms can be important (see reviews by Fender \& Gallo 2014; Zhang 2013; Neilsen 2013; Diaz Trigo \& Boirin 2013; Belloni 2010; Done et al. 2007).
Low-mass X-ray binaries (LMXBs) are especially suited for testing the theory of accretion disk winds, as the secondary in these systems is a normal star undergoing Roche Lobe overflow, and hence the blueshifted X-ray absorption lines (mostly in the \ion{Fe K}{} band) observed in some systems are attributable to a disk wind and not the stellar wind of the secondary 
[examples include GRS 1915+105 (Lee et al. 2002; Neilsen \& Lee 2009; Ueda et al. 2009); Cir X-1 (Schulz \& Brandt 2002); GRO J1655-40 (Miller et al. 2006a; Diaz Trigo et al. 2007); 4U 1630-472 (Kubota et al. 2007); H 1743-322 (Miller et al. 2006b); IGR J17091-3624 (King et al. 2012; Janiuk et al. 2015); and MAXI J1305-704 (Miller et al. 2014)].
These winds have been shown to be both massive (with wind mass loss rates as high as 20 times the accretion rate) and relatively fast (velocities of $10^2 - 10^3~\rm{km\,s^{-1}}$) [e.g., D{\'{\i}}az Trigo \& Boirin 2013; Nielsen 2013], meaning that they are very important dynamically for understanding the phenomenology of LMXBs.  For example, by removing vast quantities of matter from the disk, these winds may quench relativistic jets and contribute to state transitions (e.g., Nielsen \& Lee 2009; Ponti et al. 2012; King et al. 2013).  
The challenge is therefore to decipher which launching mechanism dominates at various distances from the compact object.

Helpful in this regard is the fact that many LMXBs do not exhibit the right conditions necessary for radiative driving (e.g., Proga \& Kallman 2002).  For example, the black hole LMXB GRO J1655-40 was found by Miller et al. (2006a) to have a luminosity only $4\%$ of Eddington, meaning that continuum driving is not important, while line driving is negligible due to the very high ionization parameter in the wind, $\log(\xi) > 4$ (the photoionization parameter $\xi$ is defined  in \S{3.4}).  
Miller et al. (2006a) further argued that thermal driving cannot be the primary launching mechanism because the wind location inferred from photoionization modeling of the observed spectrum (using a constant density slab) is just hundreds of Schwarzchild radii $R_S$, whereas thermal winds driven by X-ray irradiation cannot be launched at radii within about $0.1 R_{IC}$ (Begelman et al. 1983; Woods et al. 1996), where $R_{IC}$ is the Compton radius.  This is the characteristic distance where the Compton temperature $T_{IC}$ equals the `escape temperature', $GM_{BH}\bar{m}/k R_0$ (with $\bar{m} = \mu m_p$ the mean particle mass in relation to the proton mass, $m_p$), at distance $R_0$ along the disk:  
\beq
R_{IC} = \f{1}{2} \f{c^2}{k T_{IC}/\bar{m}} R_S = 5.45\times10^5 \mu \left(\f{T_{IC}}{10^7\rm{K}}\right)^{-1} R_S.
\label{R_IC}
\seq
Since $0.1 R_{IC}$ is at least two orders of magnitude greater than the wind location inferred by Miller et al. (2006a), they concluded that the disk wind in GRO J1655-40 must be magnetically driven.

Many different analyses have been performed to further assess the wind properties in GRO J1655-40 (e.g. Netzer 2006, Miller et al. 2008, Kallman et al. 2009; Neilsen \& Homan 2012), yet there continues to be controversy regarding the gross properties of its wind.  The early controversy centered around the density diagnostics necessary to constrain the distance of the wind through the photoionization parameter (for a summary of this issue, see Luketic et al. 2010, hereafter L10; see also Neilsen 2013).  
More recently, the sub-Eddington luminosity inferred by Miller et al. (2006a) has been challenged, again calling into question the inferred wind location.     
For example, Uttley \& Klein-Wolt (2015) suggest accretion close to or above the Eddington limit based on the timing properties observed with the \emph{Rossi X-ray Timing Explorer},
while Neilsen et al. (2016) found that the X-ray continuum is likely optically thick Compton scattering rather than the standard disc/power-law emission.  
Shidatsu et al. (2016) similarly find that a super-Eddington disk and Compton thick flow better fits the optical/UV emission, and they associate the outburst state with an Eddington luminosity >70\%.  The present state of affairs therefore renews the possibility that radiation pressure is important in GRO J1655-40.

Unfortunately, it is much more difficult to model winds in the Compton thick regime, and the present study is only applicable to sub-Eddington sources that are appropriately modeled using an optically thin approximation.  When radiation forces can be neglected, it is necessary to determine if magnetic driving is required as originally suggested by Miller et al. (2006a), or if thermal driving alone can account for the winds observed in the high-soft states of LMXBs.  L10 designed simulations to assess if thermal driving alone is indeed plausible in a multi-dimensional physical model; they found that both the density and velocity of the wind are too low to account for the observations of GRO J1655-40.  Higginbottom \& Proga (2015; hereafter HP15) explored the parameter space of the heating and cooling prescription employed by L10 and found that a denser and faster thermal wind can be obtained in instances in which more efficient heating occurs at lower $\xi$: for the same X-ray flux, launching at a lower $\xi$ corresponds to denser gas being accelerated.  
However, upon employing a more self-consistent heating and cooling prescription using rates obtained from Cloudy calculations, Higginbottom et al. (2017) found that while denser winds are indeed obtained, they are limited to velocites of only $\sim 200~\rm{km}\,s^{-1}$.  Very recently, Higginbottom et al. (2018) have further refined this model to better account for radiative transfer effects, finding slightly lower maximum velocities.  The observed lines in GRO J1655-40 are blueshifted in the $\sim 300-1600~\rm{km}\,s^{-1}$ range with a best fit radial velocity at $\sim 500~\rm{km}\,s^{-1}$ (Miller et al. 2008).  
Considering the progress made in producing nearly adequate densities via a thermal wind, in this paper we explore the possibility of obtaining the necessary velocity `boost' by including large scale magnetic fields in the thermal wind models explored by L10 and HP15.  

Previous efforts to understand thermal effects in MHD models tailored to LMXBs have used self-similar solutions (e.g., Chakravorty et al. 2016; Fukumura et al. 2017; Marcel et al. 2018).  In comparing these models with observations, the uncertainties characterizing the synthetic spectra are likely quite large because the (sometimes isothermal) temperature distributions of the solutions can be very different from the temperatures used in the separate photoionization modeling calculations.  
To improve on these efforts, we have developed time-dependent MHD models that include realistic heating and cooling rates appropriate for LMXBs using the new publicly available MHD code \athena (Stone et al., in preparation).  
Contrary to our expectation noted above, in our attempt to develop an adequate magnetothermal wind model for LMXBs, we have instead identified a circumstance in which adding magnetic fields suppresses the `primary' thermally driven disk wind.  The potential for this occurrence is easy to recognize in hindsight: a strong poloidal magnetic field can reorient the wind by imposing a streamline geometry that is nearly parallel to the magnetic field.  The new path that this flow must traverse may not be at all conducive to wind acceleration if, for example, the gravitational potential went from falling off radially to falling off only vertically, or if the flow tube geometry went from converging-diverging to diverging-converging.  We show that the cause of the suppression is essentially due to the latter.  

This paper is organized as follows.  In \S{2} we provide a bit more background into past work on MHD winds and thermal winds.  In \S{3} we describe our methods to arrive at magnetothermal wind solutions by solving the ideal MHD equations including heating and cooling.  In \S{4} we present the results of these simulations, and in \S{5} we conclude with a discussion of the implications of our results for LMXBs.  

\section{Preliminaries}
Most analytic and numerical studies on MHD winds adopt either a `cold' flow framework, valid when magnetic fields are strong enough to warrant neglecting gas pressure effects entirely, or by assuming isentropic flow, which implies neglecting all forms of heat deposition.  Detailed reviews of the basic theory under these circumstances have been presented by Spruit (1996), Tsinganos (2007), and K{\"o}nigl \& Salmeron (2011), with Ferreira (2007) summarizing the results from the first attempts to account for the effects of heat deposition.   
Pudritz et al. (2007) reviewed much of the early numerical studies on MHD winds (see also K{\"o}nigl \& Pudritz 2000).  Livio (1997) summarized basic jet scaling relations considering applications to many systems, and the recent review by Hawley et al. (2015) provides a complimentary overview of theoretical progress made in understanding the disk-jet connection.  

Analytic studies are greatly aided by the existence of four invariants that follow from the steady-state, axisymmetric, ideal MHD equations.  These represent the conservation of the flow of mass, field-line angular velocity, angular momentum, and energy along poloidal flux surfaces.  Alternatively, such flows are amenable to solution via a single partial differential equation, the Grad-Shafranov equation (e.g., Lovelace 1986).  However, it is rarely emphasized that the solutions obtained are not the most general under steady-state axisymmetric conditions.  As pointed out by Contopoulos (1996), this ideal MHD framework further requires that poloidal magnetic field lines be parallel to the poloidal velocity field.  This is the additional physical assertion that magnetic flux not be advected inward (or outward) in a steady-state and is equivalent to the requirement that the toroidal component of the electric field is zero, a condition not likely to be met in numerical simulations unless it is explicitly enforced.  When $E_\phi \neq 0$, all four of the flux/flow invariants are lost (Contopoulos 1996), but in practice they often remain approximately constant in quasi-steady flow regimes, even in non-ideal MHD simulations, and therefore serve as useful diagnostics for numerical studies.  For example, Murphy et al. (2010) find that beyond the resistive disk region, the profiles of invariants along a flux surface level off to near constant values, while Tzeferacos et al. (2013) quote deviations within $5\%$ in this region for simulations that  included magnetic resistivity and dissipative heating.  
 
Note that in much of the literature the labels `MHD jets' and `MHD disk winds' are interchangeable.  
That's mainly because the MHD equations are scale-free in the absence of source terms such as optically thin heating and cooling.  The degree of collimation or terminal velocity may serve as physical criteria for distinguishing between jets and disk winds in scale-free solutions, but in any case most past MHD `jet' studies are directly relevant to this work, even though we are clearly exploring MHD disk winds near the Compton radius and thus very far from the compact object (see equation \eqref{R_IC}).  Most notably, the early numerical studies by Ouyed \& Pudritz (1997; 1999) highlighted the importance of `mass-loading' in determining whether the outflow is steady or episodic.  They concluded that only high enough mass loads lead to steady outflow, finding that the kinetic energy at the base of the wind must sufficiently exceed the magnetic energy of 
the toroidal magnetic field component.  The opposite conclusion was reached by Anderson et al. (2005), who demonstrated the existence of a critical mass-loading rate \emph{below} which the flow is steady.  While this discrepancy appears still not fully resolved, it serves to highlight the various numerical subtleties that are required to make comparisons with steady-state MHD wind theory.  For instance, both studies used a procedure to inject material into the wind using a midplane boundary condition with a nonzero vertical velocity.  This is complicated by the fact that the nature of the boundary condition differs if the injection velocity is sub-slow magnetosonic (e.g., Ustyugova et al. 1999) or not.  We are able to avoid introducing this free parameter, as the mass load is self-consistently determined by the existence of a `primary' thermal wind (see \S{\ref{procedure}}).

An important conceptual development was made by drawing a distinction between two different types of magnetic launching (see the review by Sauty et al. 2002, for example).  
The magneto-centrifugal mechanism discussed in the introduction can be viewed as an indirect action of the Lorentz force, $\mathbf{F} = \mathbf{j}\times \mathbf{B}/c$: when the energy of the poloidal field component dominates both that of the toroidal component and the gas, the field line tension initially enforces corotation of the matter with the disk and the gas accelerates due to centrifugal force.  If instead the toroidal field dominates (requiring a strong poloidal current density, $\mathbf{j}_p$), the Lorentz force directly accelerates the gas; this is typically referred to as `magnetic pressure driving' (e.g., Uchida
\& Shibata 1985; Pudritz \& Norman 1986).  Formally, this distinction is drawn by decomposing $\mathbf{F}$ into components perpendicular and parallel to the poloidal field (Ferreira 1997),
\beq 
\begin{split}
&F_\phi = \f{B_p}{2\pi r} \nabla_\parallel I ,\\
&F_\parallel = -\f{B_\phi}{2\pi r} \nabla_\parallel I .
\end{split}
\label{LorentzForce}
\seq
Here, $I = 2\pi r B_\phi$ is the total current flowing within a given magnetic surface, and the projected gradient is defined by $\nabla_\parallel \equiv B_p^{-1}(\mathbf{B}_p \cdot \nabla)$.  Notice that the current leakage through a flux suface, $\nabla_\parallel I$, is not relevant for assessing the relative importance of these forces, as their ratio is simply $F_\phi/F_\parallel = -B_p/B_\phi$.  Since $F_\phi$ provides the torque necessary for the corotation of field lines, magneto-centrifugal launching requires $B_p >> |B_\phi|$, while winds can be driven by magnetic pressure in the opposite limit.  

The literature on thermal disk winds is much smaller than that on MHD winds in the context of accreting black hole systems.  In the specific application to Compton heated winds, studies employing time-dependent numerical simulations have been limited to those mentioned in the introduction.  In the protoplanetary disk community, there have been numerous studies on thermal winds focused on the role of disk irradiation in photoevaporating the disk (see the review by Gorti et al. 2016).  Few papers have focused on the combined effects of magnetic fields, and those that do adhere to the above framework.  For example, the recent analytic work by Bai et al. (2017) used the four MHD invariants to obtain magnetothermal wind solutions after imposing a straight fieldline (and hence streamline) geometry.  We have previously investigated the role of various straight streamline geometries in the absence of magnetic fields and showed that the flow acceleration is substantially reduced when the poloidal streamlines are parallel to each other (Waters \& Proga 2012).  The important point is that the velocity profile of the flow along the disk is sensitive to the degree of divergence of neighboring streamlines.  For purely thermal winds, the streamline geometry is mainly determined by the steepness of the midplane density profiles; substantial poloidal streamline divergence occurs for profiles steeper than $\rho \propto r^{-2}$, while nearly parallel streamlines are naturally obtained otherwise (Font et al. 2004; see also L10 and Clarke \& Alexander 2016).  As it turns out, the reduction in wind velocities that can occur in our magnetothermal wind solutions are ultimately a result of the magnetic field changing the initial streamline geometry of the primary thermal wind.

\subsection{On the use of hydrostatic disks versus disk boundary conditions}
Most studies have followed one of two basic approaches to simulate MHD or thermal disk winds.  The first approach has been to simply enforce prescribed radial profiles of the density, velocity, and pressure along the disk midplane and then to only simulate a domain from $\theta = 0^\circ-90^\circ$, applying reflecting boundary conditions (BCs) at $\theta = 90^\circ$.  We refer to this as employing `midplane BCs'.  The second approach has been to prescribe an axisymmetric hydrostatic disk solution that when evolved in the absence of any magnetic fields would closely maintain its initial conditions (ICs).  This latter approach has gained popularity in recent years, beginning with Zanni et al. (2007), who introduced a resistive magnetohydrostatic disk setup.   
We show here that these two setups lead to essentially equivalent ICs, with the usage of the disk BC simply a subgrid model for the accretion process.  That is, the static BC that we employ holds the disk midplane fixed to its initial density, velocity, and pressure profiles and therefore represents a model of a steady disk that continually replenishes any mass lost to the wind.  
Not replenishing matter lost to the disk wind in some manner would prevent the possibility of obtaining a steady state solution and therefore does not constitute a realistic simulation of a LMXB for times approaching the disk depletion timescale, $t_{\rm{dep}} \equiv M/\dot{M}$, because in reality gas from the secondary is continually being fed into the disk.  
The drawback to this approach is that we cannot self-consistently assess the role of angular momentum transport in the disk due to the wind because our boundary condition holds the midplane angular momentum constant.  

The approach developed by Zanni et al. (2007) to instead continually supply matter by way of accretion using an $\alpha$-prescription (through their inclusion of a finite magnetic resistivity) is arguably more physical, but it is less suited to drawing comparisons with steady state ideal MHD wind theory than a setup employing a midplane BC.  
Indeed, in the presence of a nonzero $\alpha$ it is actually inconsistent to impose the requirement that $\gv{B}_p$ and $\gv{v}_p$ be parallel in the disk midplane \emph{if the magnetic field there is vertical}, as Zanni et al. (2007) assume it is.  Since the angle between $\gv{B}_p$ and $\gv{v}_p$ is $\gv{B}_p \times \gv{v}_p/(B_p v_p)$, clearly $\gv{v}_p$ cannot have the radial component a nonzero $\alpha$ requires when $\gv{B}_p$ is purely vertical.  
Midplane BCs with $v_r = 0$ circumvent this issue and can therefore be viewed as an $\alpha$-prescription for steady accretion in the limit that $\alpha \rightarrow 0$, thereby providing mass replenishment in the absence of accretion.

\begin{figure}
\includegraphics[width=0.47\textwidth]{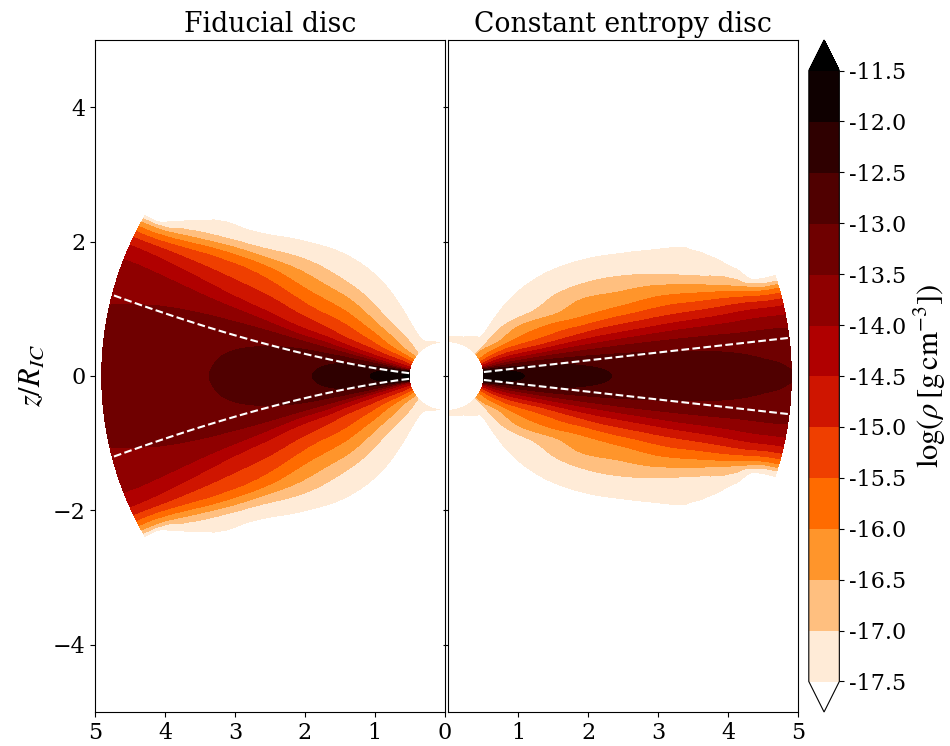}
\includegraphics[width=0.47\textwidth]{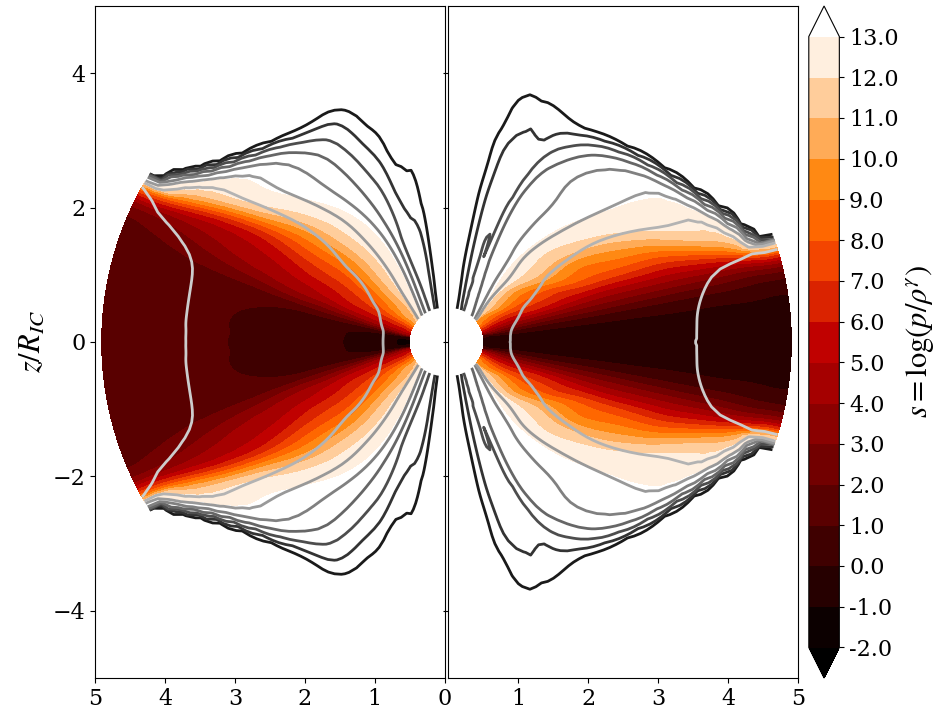}
\includegraphics[width=0.47\textwidth]{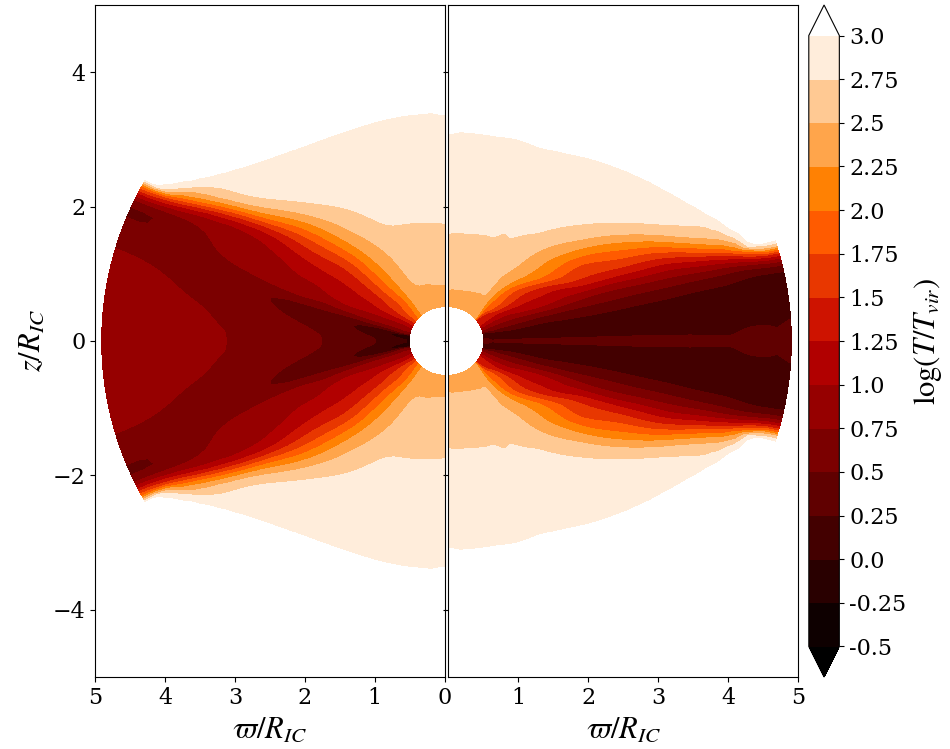}
\caption{
Color maps of density (top, in $\rm{cgs}$ units), entropy $s \equiv \log(p/\rho^\gamma)$ (middle), and temperature (bottom, in units of the local virial temperature, $T_{vir} \equiv G M^2/(3\,k\,n\,r)$) for two different hydrostatic disk solutions.  The left half of each panel shows our fiducial disk with density scaling as $r^{-2}$ and temperature constant along the midplane (therefore $T/T_{vir} \propto r$).  The right half displays a more commonly used disk with density scaling as $r^{-3/2}$, $T/T_{vir}$ constant along the midplane, and $s$ constant within about one scale height, $\epsilon \equiv c_s/v_{\rm{kep}}$ (white dashed lines in the top panels show $\epsilon(r)$).  Contours of specific angular momentum are overplotted in the middle panel with each contour decreasing by a factor of 2 relative to the lightest shade of gray.}  
\end{figure}

\subsection{A procedure to `generate' hydrostatic disks using disk boundary conditions}
\label{MBCs}
Using \athena, we demonstrate that midplane BCs yield nearly equivalent ICs to the hydrostatic disk setup employed by many authors through the following numerical experiment.  We set the density, velocity, and pressure in the first active zones above and below $\theta = 90^\circ$ (hereafter denoted the `midplane profiles') as,
\beq 
\begin{split}
\rho(r) & = \rho_0 (r/R_{IC})^{-1/(\gamma-1)},\\
v_\phi(r)  & = v_{\rm{kep}}(r)\sqrt{1- \gamma \epsilon^2/(\gamma-1)}, \\
p(r)  & =  \rho_0 v_{\rm{kep}}^2(R_{IC}) \epsilon^2 \left[\rho(r)/\rho_0\right]^\gamma.
\end{split}
\label{CED}
\seq
Here $v_{\rm{kep}}(r) = \sqrt{GM_{BH}/r}$ is the Keplerian speed and $\epsilon \equiv c_{iso}(r)/v_{\rm{kep}}(r)$ is the disk scale height parameter with $c_{iso}$ the isothermal sound speed.  Notice that when $\epsilon$ is a constant, the entropy profile will be a constant, and we must have that $c_{iso}(r) \propto r^{-1/2}$.  

For all other zones, we set $\rho = \rho_a$, $v_r = v_\theta = v_\phi = 0$, and $p = (k T_{IC}/\bar{m}) \rho_a$, where $\rho_a= 10^{-5} \rho_0$ specifies the density of a tenuous disk atmosphere.   
We then evolve these ICs to a steady state, holding the midplane profiles fixed with time.  The result for $\epsilon = 0.09$ is shown on the right hand panels of Figure~1.  

The midplane profiles in equation \eqref{CED} correspond to the 2D analytic solution of a \emph{constant entropy disk} 
evaluated at $\theta = \pi/2$.   
This one-parameter, 2D axisymmetric hydrostatic vacuum solution in spherical coordinates is
\beq 
\begin{split}
\rho(r,\theta) & =\rho_0 \left[\f{1}{a} - \f{1-a}{a\sin\theta} \right]^\f{1}{\gamma-1} (r/R_{IC})^{-1/(\gamma-1)} ,\\
v_\phi(r,\theta)  & = v_{\rm{kep}}(r) \sqrt{ \f{1-a}{\sin\theta} }, \\
p(r,\theta)  & = \f{\gamma - 1}{\gamma} \rho_0 v_{\rm{kep}}^2(R_{IC})  a \left[\f{\rho(r,\theta)}{\rho_0}\right]^\gamma,
\end{split}
\label{CEDfull}
\seq
where the governing parameter, $a \equiv \gamma \epsilon^2/(\gamma -1)$, is the enthalpy at $R_{IC}$ along the disk midplane normalized by $v_{\rm{kep}}^2(R_{IC})$.  (See Zanni \& Ferreira (2009) for the additional terms when including a nonzero resistivity.)  By `vacuum solution', we mean that there is a critical scale height above the disk at which the density and pressure vanish, 
corresponding to the critical angle $\theta_c = \sin^{-1}(1-a)$.  
Since numerical codes cannot handle vanishing densities and pressures, the atmosphere above the disk is matched to the following non-rotating ($v_\phi = 0$) spherically symmetric hydrostatic atmosphere solution, 
\beq 
\begin{split}
\rho(r,\theta) & =\rho_a (r/R_{IC})^{-1/(\gamma-1)} ,\\
p(r,\theta)  & = \f{\gamma - 1}{\gamma} \rho_a v_{\rm{kep}}^2(R_{IC}) [r/R_{IC}]^{-\gamma/(\gamma-1)}.
\end{split}
\label{CEDatm}
\seq
Pressure balance with the disk defines a disk surface at $\theta_{\rm{surf}} = \sin^{-1}[(1-a)/(1-a \rho_a/\rho_0)]$, which tends to $\theta_c$ as $\rho_a \rightarrow 0$.  
This configuration serves as the ICs in many recent numerical MHD wind papers\footnote{More accurately, equations \eqref{CEDfull}-\eqref{CEDatm} \emph{plus} the initial magnetic field serve as the ICs in the papers cited.  The point is that the numerical solution obtained by evolving these equations in the absence of any magnetic field are essentially equivalent to the solution obtained by simply evolving the midplane BCs in equation \eqref{CED} along with some low density atmosphere.  We note that an alternative commonly employed setup (see for example Zanni et al. 2007; Tzeferacos et al. 2009, 2013; Stepanovs et al. 2014; Sepanovs \& Fendt 2014, 2016) first arrives at magnetohydrostatic as opposed to just hydrostatic disk midplane profiles by including the contribution of the magnetic field to the force balance.}
(e.g., Zanni \& Ferreira 2009; Murphy et al. 2010; Sheikhnezami et al. 2012; Fendt \& Sheikhnezami 2013; Sheikhnezami \& Fendt 2015, 2018; Fendt \& Ga{\ss}mann 2018).  If we instead evolve these ICs, \emph{not} applying midplane BCs, we obtain solutions that by eye are identical to those plotted on the right hand panels of Figure~1 (quantitative differences amount to less than $5\%$ in any given variable).  Thus, we have shown that we can recover the actual \emph{numerical} hydrostatic disk solution employed in many studies simply by evolving a consistent set of midplane BCs.

\section{Methods} 
Our overall setup is a modified version of that used by L10 and HP15 (see also Proga \& Kallman 2002).  
L10 and HP15 explored domain sizes $[r_{\rm{in}},r_{\rm{out}}] = [0.05 R_{IC}, 20 R_{IC}]$ and $[0.05 R_{IC}, 2 R_{IC}]$, respectively, as Compton heated disk winds are strongest at and beyond $1R_{IC}$.  
In this work we adopt $[r_{\rm{in}},r_{\rm{out}}] = [0.5 R_{IC}, 5 R_{IC}]$.  
A more distant inner boundary has been chosen because the region within $0.5 R_{IC}$ hosts a highly bound atmosphere that becomes very turbulent upon threading it with magnetic field lines, and our default resolution becomes insufficient to capture its dynamics.  This issue can be overcome by using adaptive mesh refinement, which is currently available in \athena, but we leave this for future work as our focus is on the properties of the outer disk wind.  The choice of outer radius at $5 R_{IC}$ was made simply to give a sufficient dynamical range ($r_{\rm{out}}/r_{\rm{in}} = 10$ is still quite small for a global simulation).  

\subsection{Fiducial disk vs. constant entropy disk}
We present solutions using two different hydrostatic disks as initial conditions.  
Because numerous past MHD studies use the `constant entropy disk' from \S{\ref{MBCs}}, we consider this case in addition
to the one commonly employed in thermally driven disk wind studies (L10; HP15; Higginbottom et al. 2017) --- our `fiducial' disk --- which has a midplane density profile $\rho(r) \propto r^{-2}$.
The left hand panels of Figure~1 show our fiducial disk, obtained by the same procedure described in \S{\ref{MBCs}},
but instead of evolving the midplane profiles in equation \eqref{CED} we use 
\beq 
\begin{split}
\rho(r) & = \rho_0 (r/R_{IC})^{-q},\\
v_\phi(r)  & = v_{\rm{kep}}(r)\sqrt{1-q \epsilon^2(r)}, \\
p(r)  & =  \rho(r) c_0^2,
\end{split}
\label{FID}
\seq
with $c_0 = \sqrt{k T_{IC}/\bar{m}}$ the isothermal sound speed at $R_{IC}$ and $\epsilon(r) \equiv c_0/v_{\rm{kep}}(r)$, consistent with the previous definition.  
Both equations \eqref{CED} and \eqref{FID} satisfy the radial force balance 
equation for an axisymmetric equilibrium disk,
\beq
\f{v_\phi^2}{r} = \f{1}{\rho}\f{dp}{dr} + \f{d\Phi}{dr}
\label{RFB}
\seq
for gravitational potential $\Phi = - G M/r$.  
Compared with the constant entropy disk, our fiducial disk (with $q=2$) has an increasing scale height $z_{\rm{surf}} = \epsilon \varpi$ owing to the constant temperature profile along the midplane; this scale height is marked with the white dashed line in the top panels of Figure~1.  Notice from the middle panels that the entropy is indeed constant within this scale height on the right hand panel, while it scales as $r^{-1}$ in our fiducial disk.  The main difference between these disks is their virial temperatures $T_{vir} = G M^2/(3\,k\,n\,r)$ (bottom panels), with our fiducial disk having a substantially larger $T/T_{vir}$ indicating that the gas in this disk is more loosely bound.  The consequence of this is that both the thermal and magnetothermal wind solutions from our fiducial disk have higher mass fluxes through $r_{\rm{out}}$ despite the steeper midplane density profile (see \S{4.1}).

By accounting for the sub-Keplerian rotation profile and using a domain extending from 0 to $\pi$, our thermal wind solutions are mildly different from those of L10 and HP15.  We have also run simulations with $v_\phi(r,\pi/2) = v_{\rm{kep}}(r)$ as they did and noticed only slight deviations in the global MHD disk wind properties from those reported here.  Aside from gaining consistency with the constant entropy disk, the advantage of evolving midplane profiles that satisfy equation \eqref{RFB} is a noticeably less disruptive transient upon magnetizing the steady state thermal wind solutions. 

\subsection{Magnetic fields}
Our magnetothermal wind solutions are obtained by re-evolving steady thermal wind solutions with one of two magnetic fields added: (i) a simple vertical field and (ii) the popular `Zanni-field'.  In spherical coordinates, either field is specified using a vector potential $\mathbf{A}$ with components $(0,0,A_\phi)$.  The magnetic field lines are the contours of the flux function $\Psi = A_\phi r \sin\theta$.  

A constant vertical field is given by 
\beq
A_\phi(r,\theta) = \f{B_i}{2} r\sin\theta,
\seq 
where $B_i$ is the poloidal field strength.  The Zanni-field is a purely poloidal field with a parameter $m$ controlling the initial degree of field line bending:
\begin{equation}
A_\phi(r,\theta) = \f{4}{3} B_i R_{IC} \sin\theta \left(\f{r}{R_{IC}}\right)^{-\f{1}{4}} 
\left(\f{m^2}{1 + (m^2-1)\sin^2\theta}\right)^{\f{5}{8}}.
\label{eq:Aphi_zanni}
\end{equation}
In this case, the parameter $B_i$ specifies the field strength at $R_{IC}$.  This field gives an Alfv\'en speed $v_A \propto r^{-1/2}$ along the midplane when combined with density profile scaling as $r^{-3/2}$, thereby satisfying the self-similar requirement of Zanni et al. (2007) that all characteristic speeds scale as $r^{-1/2}$ for the constant entropy disk.   
The long term evolution is found to only mildly depend on the field line bending parameter $m$ (Stepanovs \& Fendt 2014), so we adopt the commonly used value $m = 0.4$, which satisfies the Blandford \& Payne criterion for magnetocentrifugal acceleration (Tzeferacos et al. 2009). 

These magnetic fields are applied to the entire domain.
We have also explored test runs where we only magnetize regions of the domain where there are field lines with footpoints anchored in the disk, i.e. $A_\phi$ is set to zero in the polar regions within the innermost field line.  For the vertical field this region is $\varpi < \varpi_i$, where $\varpi$ is the cylindrical radial coordinate and $\varpi_i$ is the inner disk radius.  Denoting the innermost field line as $\Psi_i$ for the Zanni field and noting that $\Psi = A_\phi(r,\theta) \varpi$, this region is determined instead by $\varpi < \Psi_i/A_\phi(r,\theta)$, where $\Psi_i$ evaluates to $(4/3) B_i R_{IC} (\varpi_i/R_{IC})^{3/4}$.
These test runs show qualitatively similar results as those reported on here, but the initial transient is much more pronounced, making it more difficult to draw comparisons with steady state MHD theory.  Presumably, there is little mention of this procedure being performed in past studies due to similar reasons.   

\subsection{MRI suppression and disk shielding}
\label{MRIandshielding}
The gas temperatures of geometrically thin, optically thick accretion disks are regulated by internal radiative processes different than those of their atmospheres or disk winds, which are mainly determined by the external radiation field.  While the ultimate goal is to solve the equations of radiation magnetohydrodynamics accounting for all of these processes self-consistently, this will necessarily require having enough resolution to resolve the MRI turbulence within the disk, which in turn requires fully 3D global simulations.  Efforts to accomplish this feat have been undertaken for AGN disks (e.g., Jiang et al. 2017), but our goal here is mainly to assess if magnetothermal wind solutions can provide the necessary velocity boost needed to account for observations of systems such as GRO~J1655-40.  We therefore adopt a highly simplified treatment of the disk physics by purposefully under-resolving the MRI and then shielding the disk from the irradiation to avoid having to introduce a separate heating and cooling prescription inside the disk.  

We find that the MRI will be suppressed if the fastest growing mode is not sufficiently resolved.  
Thus we merely construct a grid with $\Delta z > \lambda_c$ in the disk, where
\beq
\lambda_c = \f{2\pi}{\sqrt{3}} \f{v_A}{\Omega} 
\seq
is the wavelength of the fastest growing MRI mode, with $\Omega = v_\phi/r$ the angular velocity.
This criterion is unsatisfied even for a relatively low 2D grid resolution of $(N_r,N_\theta) = (100,100)$.  
Therefore, we utilize the custom grid feature of \athena to design a grid that has higher resolution at the surface of the disk where the outflow is launched, but low enough resolution within the disk so that the above criterion is met.  
This grid is plotted in Figure~2, which shows that we also added increased resolution at the poles and applied a smooth gradient in the $\theta$-direction to transition between the coarse and refined regions.  The refined regions have an effective resolution $\Delta \theta = 1/160$, the coarser regions having $\Delta \theta  = 1/40$.  The radial grid has $N_r = 100$ and uses logarithmic spacing with $dr_{i+1}/dr_i = 1.01$.  

\begin{figure}
\includegraphics[width=0.45\textwidth]{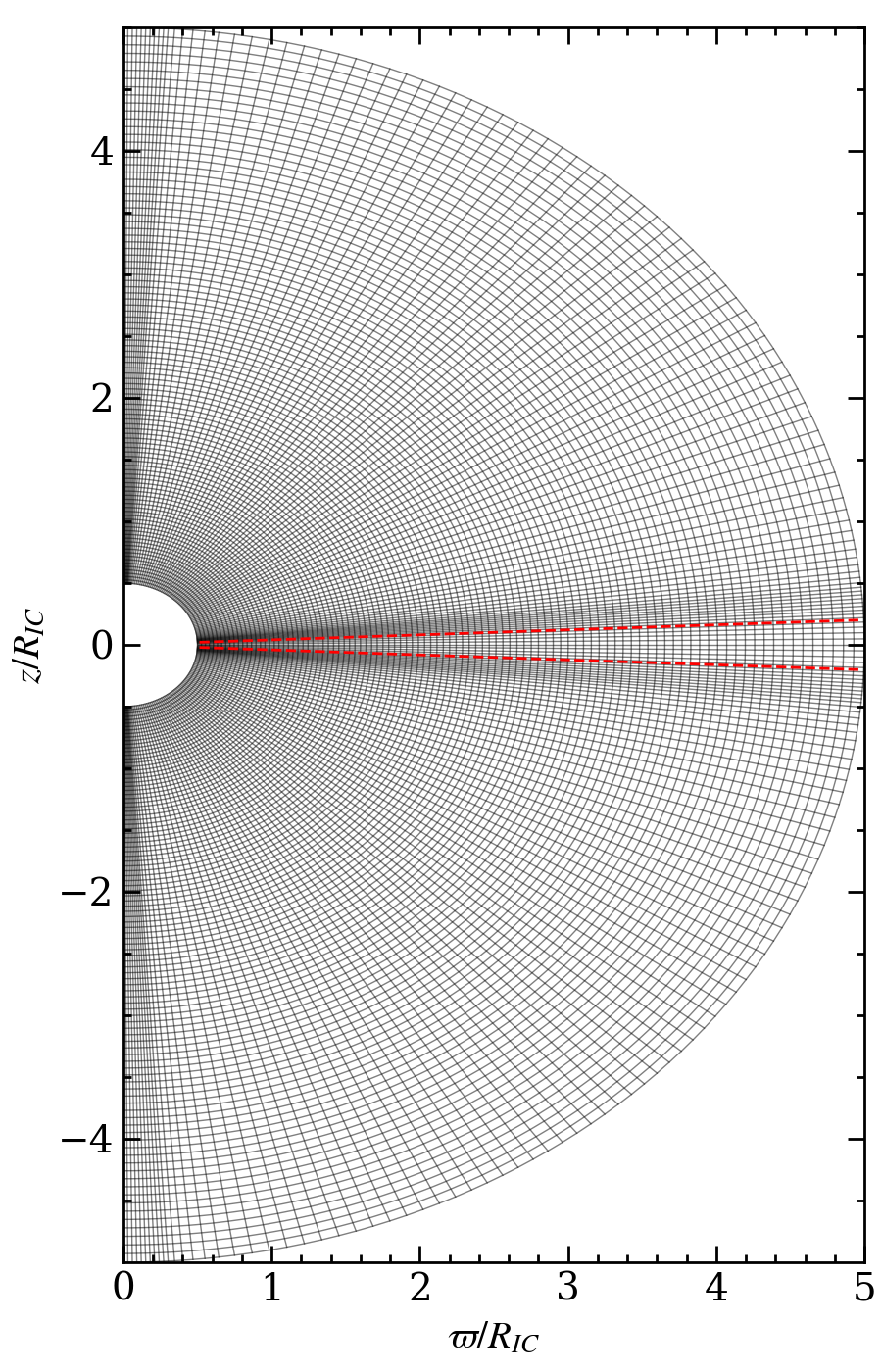}
\caption{Our computational grid.  Interior to the red dashed lines the disk is shielded from X-ray irradiation.  We utilized the custom grid capability of \athena to coarsen the grid inside this region enough so that the MRI is unresolved, ensuring that the disk serves as simply a reservoir of matter.}
\end{figure}

Our model assumes that a very compact source of X-rays irradiates the disk surface layers.  
The Compton thick regions of the disk will be shielded from these X-rays. 
Since we do not specify the properties of the disk within $0.5\,R_{IC}$, the height at which disk layers become Compton thin is a free parameter.  This height will determine the boundary between the cold disk and the Compton heated thermal disk wind.  Approximating the X-ray emission as emanating from a point source at $r=0$, this irradiation scale height $\epsilon_{wind} = z_{wind}/\varpi$ will be a constant.  We chose $\epsilon_{wind} = 0.05$ to be well within the disk scale height; this choice is marked with the dashed red lines in Figure~2.  
We implement disk shielding by setting the net cooling to zero if $|z| < \epsilon_{wind} \varpi$.

\begin{figure*}
\includegraphics[width=0.8\textwidth]{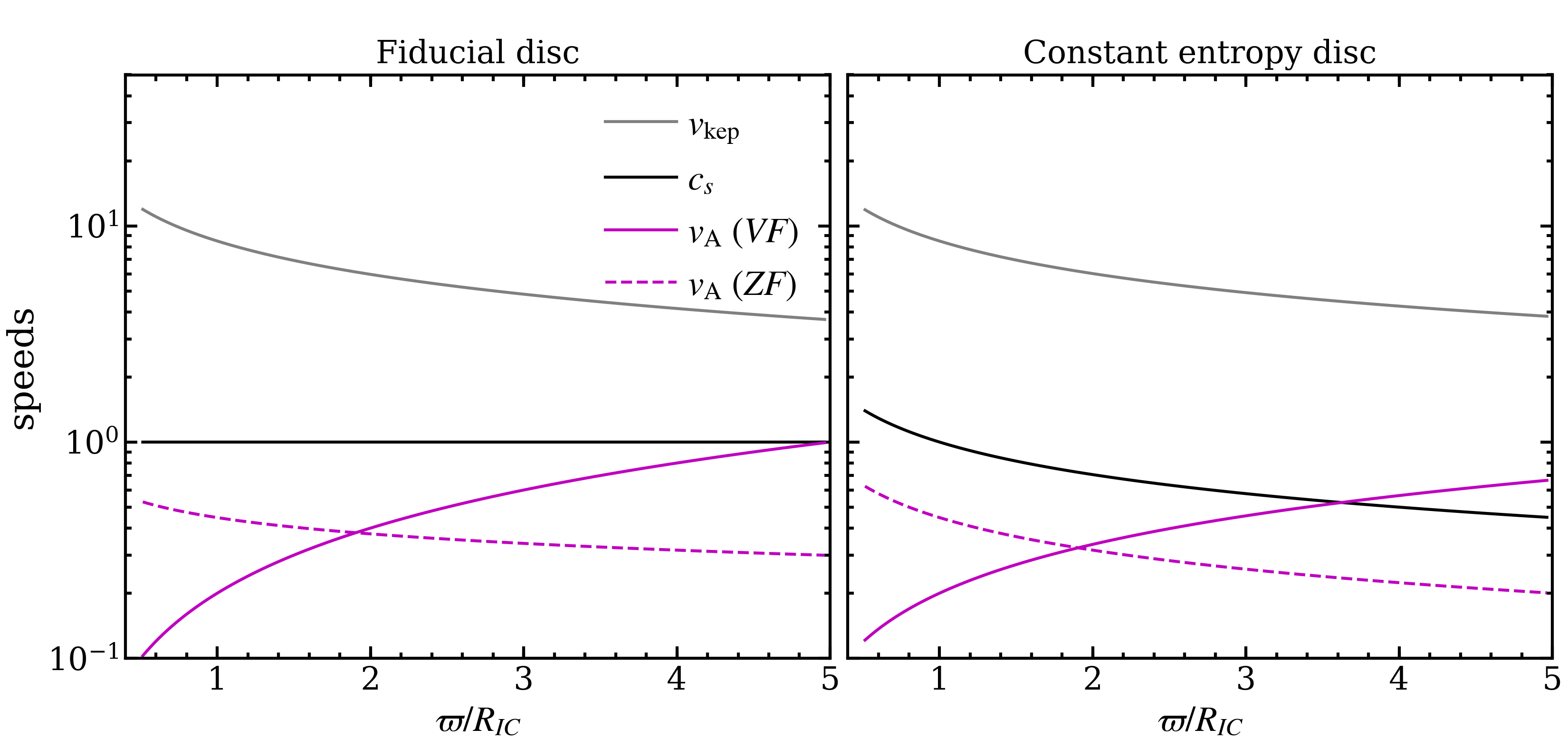}
\caption{Comparison of relevant velocities along the midplane: keplerian velocity (top gray lines), sound speed (black lines), and Alfv\'enic speed for each magnetic field (magenta lines).  These control the global properties of the resulting magnetothermal wind solutions.  We analyze four cases in total, comprised of two different magnetic field configurations (VF and ZF denoting a vertical field and the Zanni-field, respectively) each applied to steady disk winds launched from the two different disks displayed in Figure~1.    
For the constant entropy disk (right panel), notice that $v_{\rm{kep}}$, $c_s$, and $v_{A}$ all vary as $r^{-1/2}$ along the midplane in the case of a Zanni-field.  In contrast, the Alfv\'enic speed increases with radius for both disks with a vertical field.  The sound speed is constant along the midplane in our fiducial disk.}
\end{figure*}

\subsection{Solution procedure}
\label{procedure}
We solve the equations of ideal MHD with a source term added to include radiative heating and cooling processes,
\beq
   \frac{D\rho}{Dt} + \rho \nabla \cdot {\bf v} = 0,
\seq
\beq
   \rho \frac{D{\bf v}}{Dt} = - \nabla p - \rho \nabla \Phi+  \frac{1}{4\pi} {\bf (\nabla \times B) \times B},
\seq
\beq
   \rho \frac{D\mathcal{E}}{Dt} = -p \nabla \cdot {\bf v} - \rho \mathcal{L},
\seq
\beq
\label{eqn:induction}
{\partial{\bf B}\over\partial t} = {\bf\nabla\times}({\bf v\times B}).
\seq
Here, ($\rho$, $p$, ${\bf v}$, $\mathcal{E}$) are the gas mass density, pressure, velocity, and internal energy,  
respectively, 
$\Phi$ is the gravitational potential, $\bf B$ is the magnetic field vector,
and $\mathcal{L}$ is the net cooling function.
We adopt an adiabatic equation of state
$p=(\gamma-1)\rho\mathcal{E}$ and consider models with $\gamma=5/3$.
Our calculations are performed in spherical polar coordinates
$(r,\theta,\phi)$, while it is convenient to also reference cylindrical coordinates, $(\varpi,z,\phi)$.   
We assume axial symmetry about the rotational axis of the disk ($\theta=0^\circ$).  
Note that \athena solves the conservation form of equations (11)-(14).

The net cooling function $\mathcal{L} = \mathcal{L}(T,\xi)$ we use was first formulated by Blondin (1995) and has been applied extensively by our group in recent years to study the dynamics of X-ray irradiated plasmas in various contexts (e.g., Proga et al. 2000; Proga \& Kallman 2002; Kurosawa \& Proga 2009; L10; Moscibrodzka \& Proga 2013; HP15; Proga \& Waters 2015).  It assumes a $10\,\rm{keV}$  bremsstrahlung ionizing spectrum and accounts for both photoionization heating and line cooling using fits to \textsc{XSTAR} calculations first performed by Blondin (1995) and later independently confirmed by Dorodnitsyn et al. (2008).  Also included are analytic rates for Compton heating and cooling and bremsstrahlung cooling.  All of these rates are parametrized in terms of the gas temperature $T$ and photoionization parameter, $\xi$, which is defined in terms of hydrogen number density $n_H$ as
\beq
\xi = \f{4\pi F_X}{n_H} = \f{L_* }{r^2 n_H}e^{-\tau},
\seq
where $F_X$ is the local X-ray flux seen by the plasma.  This will be attenuated relative to the flux assigned to the inner boundary of the computational domain, $F_* = L_*/4\pi r_{\rm{in}}^2$, with $L_*$ the luminosity of the X-ray source.  As in HP15, we adopt a luminosity appropriate for GRO~J1655-40, namely 3.7\% of the Eddington luminosity, $L_* = 3.3\times10^{37}\rm{erg\,s^{-1}}$.  Our neglect of the radiation force is therefore justified.  The attenuation due to electron scattering is accounted for by recomputing the optical depth at every timestep using 
\beq
\tau(r,\theta) = \int_{r_{\rm{in}}}^{r} n_e(r,\theta) \sigma_e dr,
\seq
where $n_e$ is the electron number density and $\sigma_e$ the Thomson cross section.  Following Townsend (2009), we properly account for the gas mixture due to different elemental abundances by tracking $n_H = \rho/(\mu_H m_p)$ and $n_e = \rho/(\mu_e m_p)$.  However, for simplicity we set $\mu_H = \mu_e = 1$ since we have not yet attempted a realistic modeling of GRO~J1655-40.  
 
The free parameters of the model are $M_{BH}$, $T_{IC}$, $\mu$, $L_*$, $\epsilon_{wind}$, $B_i$, and $\xi_0$ (the value of $\xi$ at $R_{IC}$, neglecting $\tau$).  Together, $M_{BH}$, $T_{IC}$, and $\mu$ determine $R_{IC}$, while in the photoionization parameter framework, $R_{IC}$, $L_*$ and $\xi_0$ determine the disk density at $R_{IC}$, $n_0$.  The disk temperature there, $T_0$, follows from assuming radiative equilibrium, $\mathcal{L}(T_0,\xi_0) = 0$.   
As in L10 and HP15, we set $\xi_0 = 10^{2.1}$, giving $T_0 = 1.14\times10^5\,\rm{K}$, as well as $T_{IC} = 1.4\times10^7\,\rm{K}$, giving $R_{IC} = 4.8 \times 10^{11}\,\rm{cm}$ for $M_{BH} = 7\,M_\odot$ and $\mu = 0.6$.  By the definition of $R_{IC}$, note the relation $T_0 = \epsilon^2(R_{IC}) T_{IC}$, i.e. $\epsilon(R_{IC}) = 0.09$.  See below for the values of $B_i$.

As described in \S{2.1}, we first solve these equations in the absence of magnetic fields and with $\mathcal{L} = 0$ to obtain the hydrostatic disk solutions depicted in Figure~1.  We next turn on heating/cooling, which requires additional special treatment than in previous works that used a domain extending only to $90^\circ$, as discussed above in \S{3.3}.  The global magnetic fields from \S{3.2} are then applied to the steady state thermal wind solutions, and these are the initial conditions for our MHD runs.  Our boundary conditions and numerical floor treatment are documented in the Appendix.

\begin{figure*}
\includegraphics[width=0.95\textwidth]{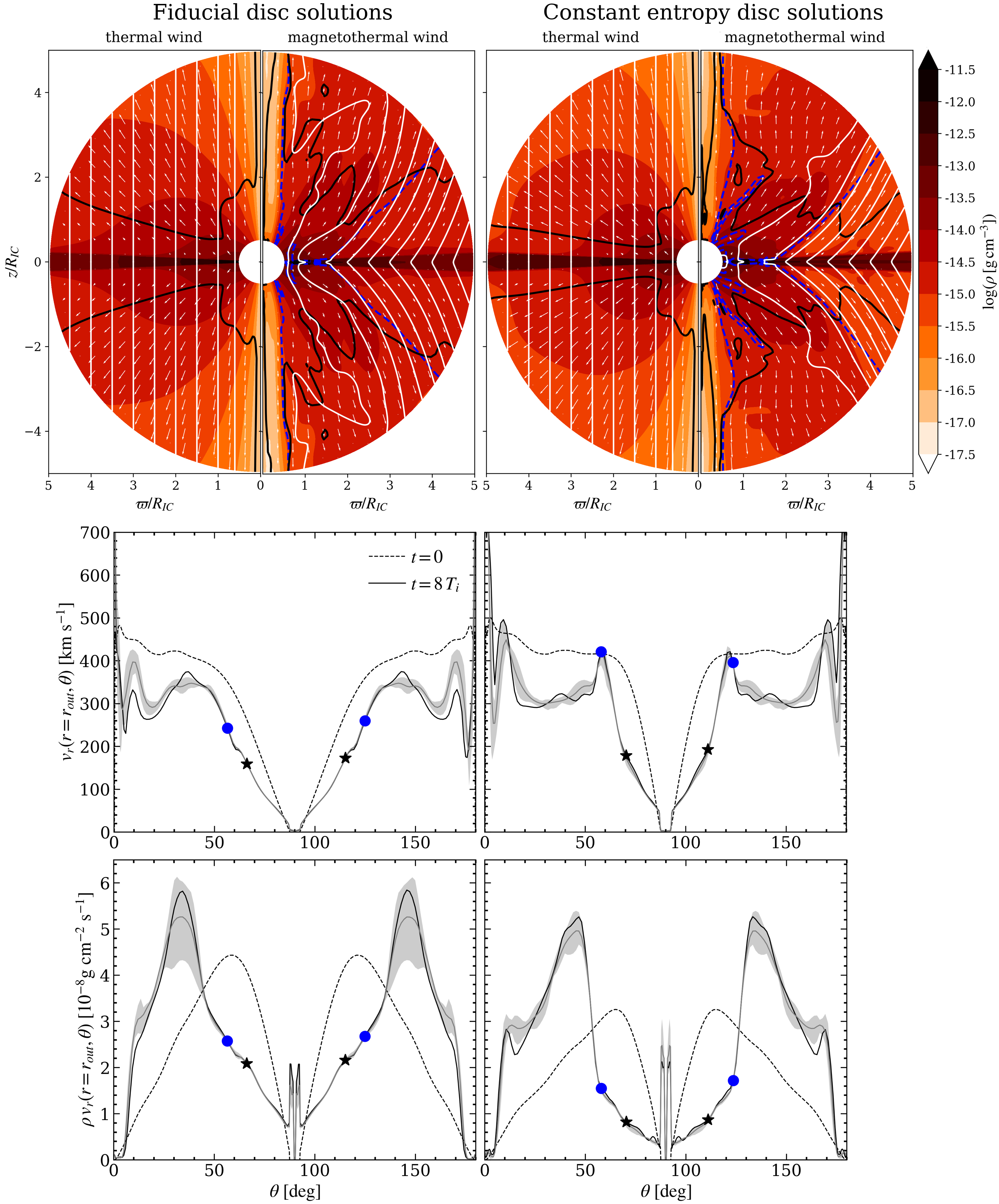}
\caption{
Global properties of vertical field runs demonstrating suppression of the thermal disk wind.  Top row: contour maps of density.  The left half of each panel shows the steady state thermal wind solutions with the initial magnetic fields (white contours) and the sonic surface (black contour) overplotted.  This serves as the initial state for the magnetothermal wind solution, which is shown on the right half of each panel after evolving the initial state for eight inner orbits.  Magnetic field lines and the sonic surface are again overplotted in addition to the Alfv\'en surface (dashed blue contour).  White arrows denote the velocity fields.  Middle row: comparison of radial velocities at $r_{\rm{out}}$ for the steady thermal wind (dashed line) and the magnetothermal wind (solid line) for each disk.  Time-variability is assessed by plotting the 25-75 percentile range (gray bands) of over 100 dumps (spanning 2 inner orbits) in the time range shown in gray in Figure~6; the gray solid line shows the mean value.  Blue circles (black stars) mark locations of the Alfv\'en (sonic) surface at $r_{\rm{out}}$.  Bottom row: same as the middle row but showing mass flux density along $r_{\rm{out}}$ instead.  
}
\end{figure*}

\begin{figure*}
\includegraphics[width=0.95\textwidth]{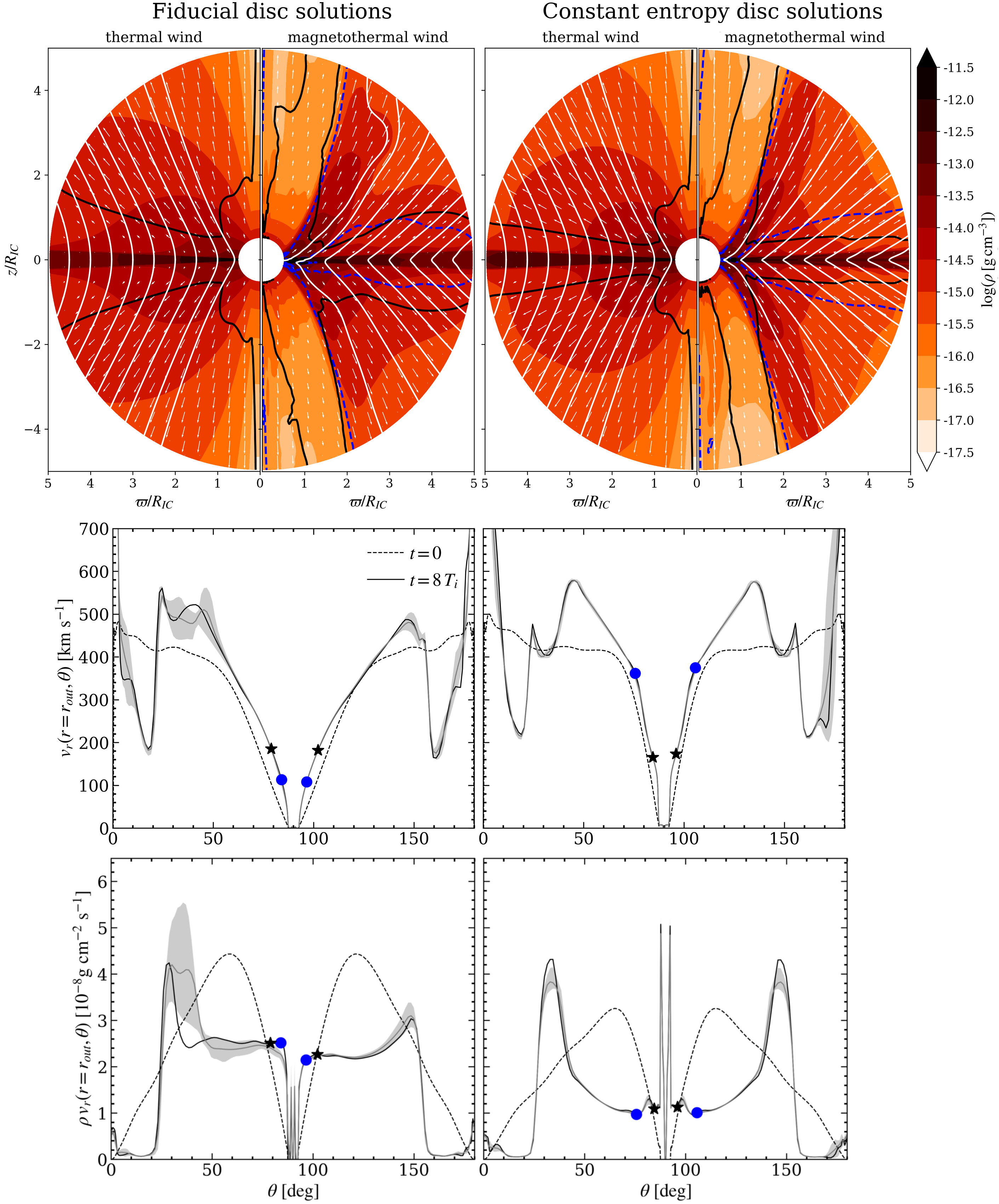}
\caption{
Same as Figure~4 but for the Zanni-field runs.
}
\end{figure*}

\section{Results}
We have examined two magnetic field configurations in detail by running dozens of simulations for each one to explore the sensitivity of the results to the field strength and geometry, as well as to the initial conditions.  Here we report on four runs that capture the essence of this sensitivity.
In line with expectations, the inclusion of magnetic fields significantly changes our thermal wind solutions only for field strengths close to equipartition with the thermal energy, which will be the case when the Alfv\'en speed, $v_A \equiv B/\sqrt{4\pi\rho}$, is not much less than the sound speed at the disk midplane.  A basic upper limit on the magnetic field strength is set by the requirement that the disk can support the magnetic field (e.g., Spruit 1996), i.e. that the magnetic energy $B^2/8\pi$ be less than the rotational energy of the disk, $\rho v_\phi^2/2$.  With $v_\phi \approx v_{\rm{kep}} = \sqrt{GM/r}$, this is equivalent to the requirement that $v_A < v_{\rm{kep}}$ along the disk midplane.  In Figure~3, we plot these relevant speeds for the two magnetic fields.   
The `Zanni-field' with $B \propto r^{-5/4}$ is strongest in the inner regions of the disk and $B_i \approx 9\,\rm{G}$ is chosen so that the poloidal plasma beta, $\beta \equiv p/(B_p^2/8\pi)$, equals 6.0 in the disk midplane at $R_{IC}$ (for both disks), corresponding to a field strength falling from $20\,\rm{G}$ at $r_{\rm{in}}$ to about $1\,\rm{G}$ at $r_{\rm{out}}$.  The vertical field has a constant field strength of $B_p = B_i \approx 4\,\rm{G}$, corresponding to $\beta = 30$ at $R_{IC}$.  These fields thus provide quite contrasting cases for exploring magnetothermal effects.

\subsection{Vertical field solutions}
The top row in Figure~4 displays density contour maps and the velocity field of our thermal (left panels) and magnetothermal (right panels) wind solutions.  
The thermal wind solutions are steady state and very similar to those presented in L10 and HP15, 
despite our somewhat different numerical setup (see \S{3.1}).  
The black contour marks the sonic surface.  Contours of the initial vertical magnetic field applied are overplotted in white, while the dashed blue contour denotes the Alfv\'en surface.  
The magnetothermal wind solutions are shown after being evolved for eight orbital times at $r_{\rm{in}}$, which is a sufficient time for the outer disk wind to reach a quasi steady-state.     
Magnetic field lines at the same footpoint locations are again overplotted along with the sonic and Alfv\'enic surfaces.  The field line behavior is a good indicator of the dual unsteady-steady nature of the disk wind.  
The flow with streamline footpoints within $\varpi = 1.5 R_{IC}$ is highly variable, despite the velocity field being super-fast magnetosonic (this critical surface is shown in Figure~7).  The time dependence is especially prominent for the fiducial disk (left panel), as high density blobs are seen propagating outward, their inertia twisting the fields as they do so.  As discussed in \S{2}, this episodic behavior should not come as a surprise considering the sensitivity of the steady nature of MHD disk winds to various mass loading prescriptions.

We compare the bulk outflow properties of the thermal and magnetothermal solutions in the middle and bottom rows of Figure~4, which show radial velocities and mass fluxes as a function of $\theta$ at $r_{\rm{out}}$ in the top and bottom subpanels, respectively.  
Our main result is the prominent reduction in the radial wind velocity, as is made evident by comparing the profiles for the magnetothermal wind solutions (black solid lines) with the thermal wind solutions (dashed lines).  The mass flux density suffers an even greater reduction $35^\circ$ above and below the disk in the steady wind region, implying a decreased wind density, which is apparent from the density maps.  Notice the steady region of the constant entropy disk has a significantly lower density compared with the fiducial disk, a consequence of the latter having a higher virial temperature as pointed out in \S{3.1}.  The prominent spikes in mass flux density in the unsteady wind region is due to an increase in density in the region $0^\circ < \theta \lesssim 55^\circ$.  However, as we show in \S{4.3}, the overall kinetic luminosity of the wind is less than that of the thermal wind.

To gauge the dual unsteady-steady nature of the disk wind we have gathered statistics for over 100 dumps centered around $t = 8\,P_i$, with $P_i = 24\,\rm{ks}$ (6.8 hours) denoting the orbital period at $r_{\rm{in}}$, and spaced $\Delta t = 0.02\, P_i$ apart.  The gray bands show the interquartile range of both $v_r$ (middle row) and $\rho v_r$ (bottom row), with the gray solid line being the mean value.  Clearly, the regions from about $50^\circ - 130^\circ$ for the fiducial disk and $60^\circ - 120^\circ$ for the constant entropy disk have reached steady state values, while the remaining region is unsteady with characteristic velocity fluctuations around $40~\rm{km\,s^{-1}}$.  The steady wind regions are sub-Alfv\'enic and highly magnetized, with $\beta << 1$ (see Figure~7).  

\subsection{Zanni-field solutions}
Adiabatic MHD solutions obtained by adding the Zanni-field 
to the hydrostatic disk plus atmosphere setup shown in the right panels of Figure 1 have been extensively studied.   
Globally quasi-steady outflows are typically obtained.  Interestingly, despite our instead applying this field to a steady state thermal wind solution that takes into account heating and cooling, the constant entropy disk with the Zanni-field is the only run (out of all four runs) that is closely quasi-steady everywhere.  These results are therefore consistent with past studies, but we find qualitatively different results with the fiducial disk.  Namely, as with the vertical field runs, these magnetothermal wind solutions are best characterized as consisting of an inner unsteady and an outer quasi-steady disk wind.  The gross properties of both solutions are presented in Figure~5.

In either case, we again see that the flow consists of a super-Alfv\'enic outflow extending from about $10^\circ$ to $60^\circ$ along the outer boundary, but the peak velocities exceed those of the thermal wind by $100$ and $200~\rm{km\,s^{-1}}$, for the left and right panels, respectively.  This is followed by a sub-Alfv\'enic disk wind that is mildly faster than the thermal wind.   
Also consistent with previous MHD wind studies using the Zanni-field (e.g., Tzeferacos et al. 2009; Murphy et al. 2010) is that wind launching points up until about $3R_{IC}$ ($4R_{IC}$ for our fiducial disk) along the midplane always become super-Alfv\'enic, while the remaining region crosses only the slow magnetosonic surface (not shown, but see Figure~8). 

An interesting property of these magnetothermal solutions is the substantial reduction of the wind density in the region $40^\circ \lesssim \theta \lesssim 75^\circ$ and $105^\circ \lesssim \theta \lesssim 140^\circ$ compared with the thermal wind solutions, as revealed by the bottom panels in Figure~5 (but also apparent in the density colormaps).  Despite the higher velocities, this leads to a significant decrease in mass flux in the outer wind region, the spike in the mass flux occupying only a small solid angle.  Thus, we can claim that the Zanni-field also suppresses the thermal disk wind in terms of mass loss rate. 

\begin{figure*}
\includegraphics[width=0.86\textwidth]{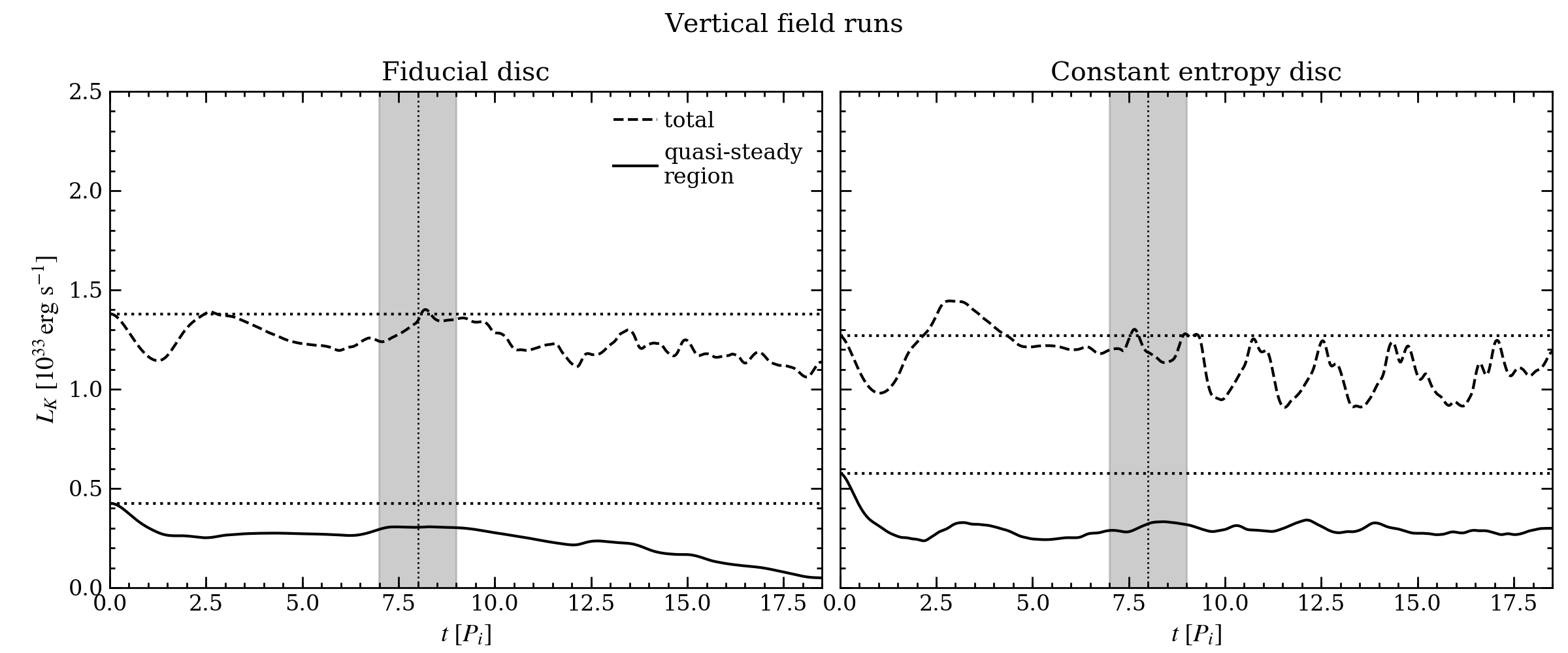}
\includegraphics[width=0.86\textwidth]{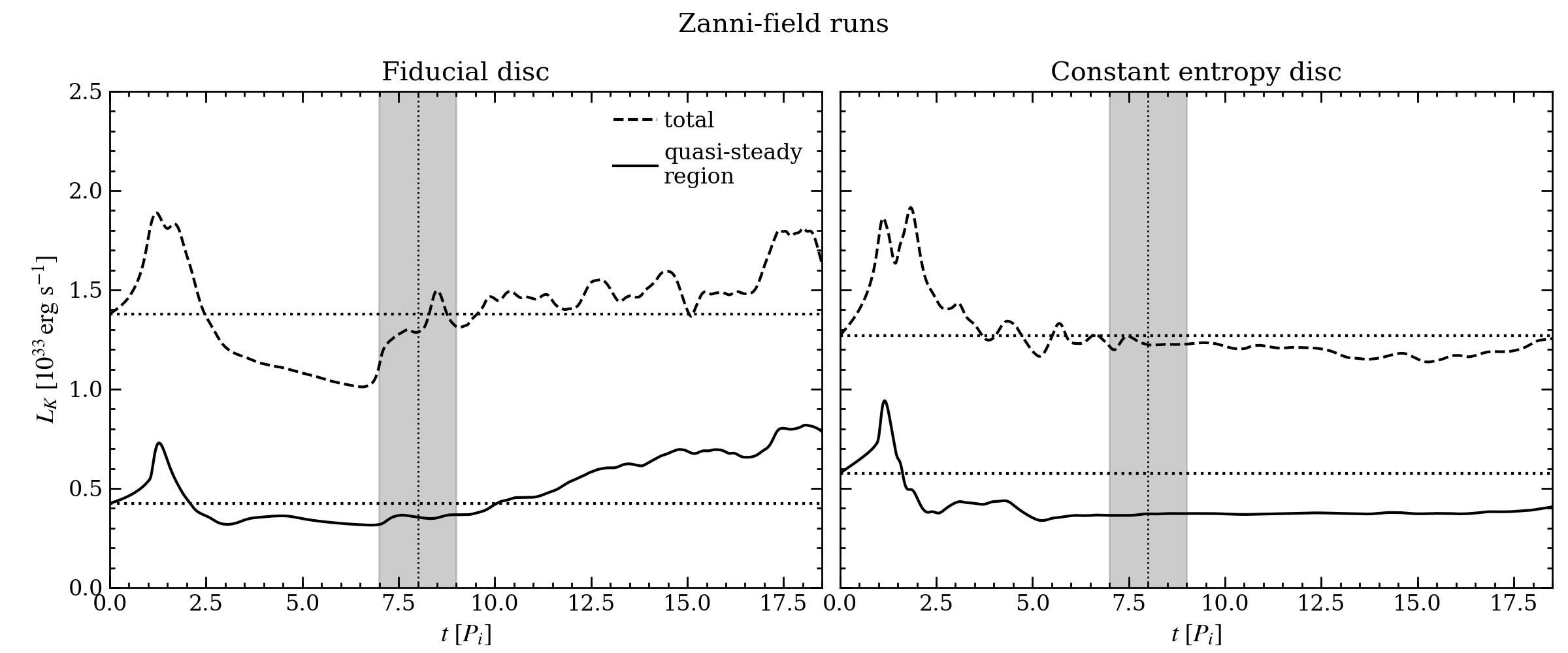}
\caption{
The kinetic luminosity defined in equation \eqref{eq:Pk} as a function of time for all of our runs.  Dashed curves show $L_K$ computed over the whole domain.  Solid curves denote the quasi-steady wind regions of these solutions, defined as the portion outflowing through angles $30^\circ$ above and below the midplane at $r_{\rm{out}}$.  Horizontal dotted lines draw a comparison with the steady thermal wind solutions.  The shaded regions represent the times used for the statistics shown in Figures~4 and 5, while the dotted vertical lines mark the times corresponding to the colormaps plotted in Figures~4,\,5,\,7 and 8.}
\end{figure*}

To support this claim we assess whether or not the associated kinetic power is overall smaller or larger than that of the thermal wind and of the vertical field solutions.  In Figure~6 we examine the time-dependence of $L_K$, defined as the flux of kinetic energy integrated over the outer boundary:
\begin{equation}
L_K = 2\pi r_{\rm{out}}^2 \int_{\theta_1}^{\theta_2} \f{\rho v^2}{2}\, v_r\sin\theta \, d\theta.
\label{eq:Pk}
\end{equation}
The dashed curves, representing the total $L_K$, reveal initial large fluctuations in $L_K$ characterizing the transient response of the solutions upon the addition of the magnetic field.  Further fluctuations are indicative of episodic ejections of disk material.  
The wind region $30^\circ$ above and below the disk midplane is quasi-steady in all of our runs (at least up until $t = 8 P_i$), as seen by calculating $L_K$ only in this region, shown as the solid curves in Figure~6. 
In all cases, the kinetic luminosity is reduced relative to that of the steady thermal wind solutions (shown as horizontal dotted lines) up until $t = 8 P_i$.  
Beyond this time, the solutions begin to suffer from artificial collimation of the magnetic field (see the Appendix), so we do not report on the subsequent behavior shown in Figure~6.  

\begin{figure*}
\includegraphics[width=0.9\textwidth]{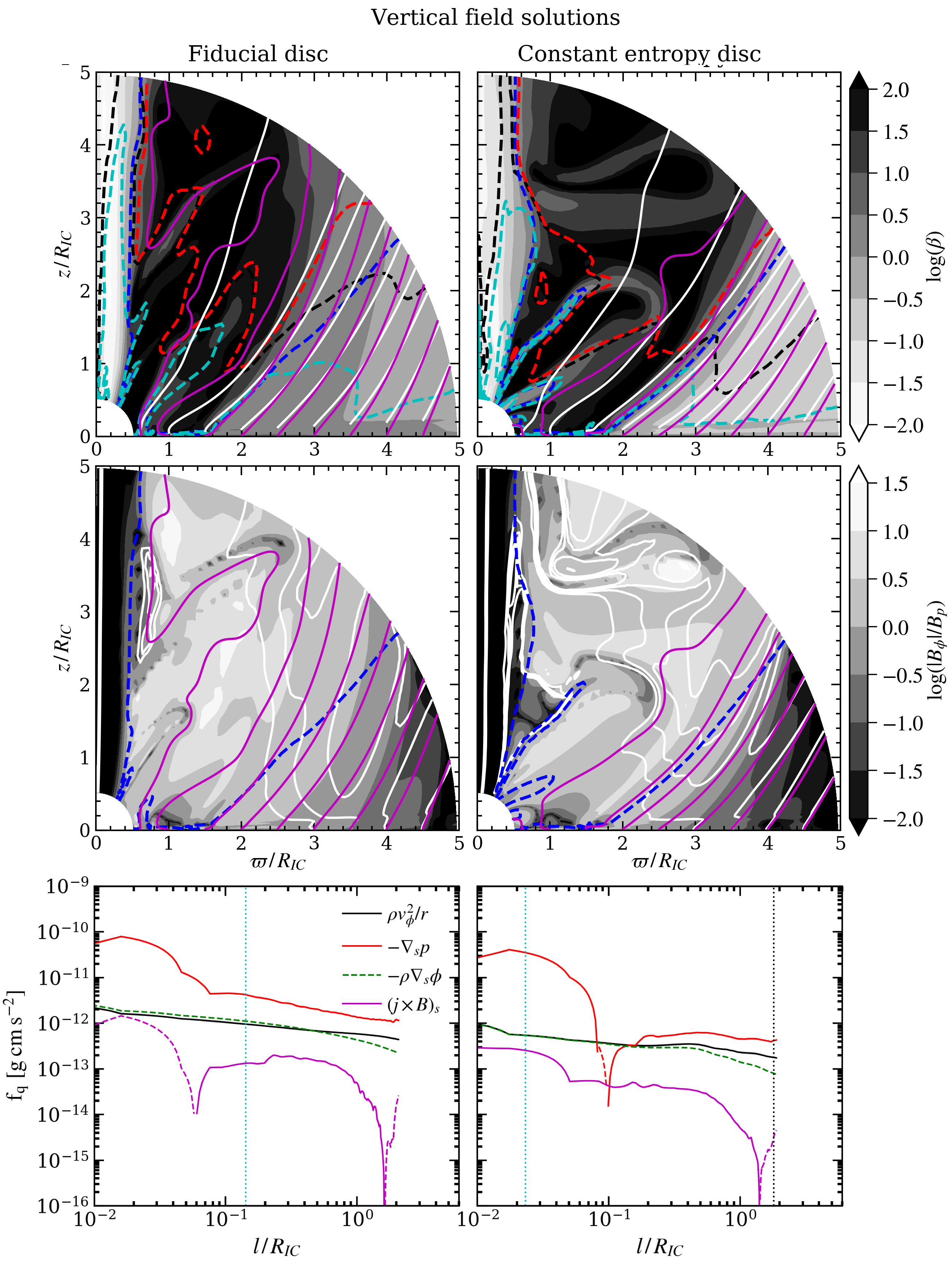}
\caption{
MHD diagnostics for the vertical field solutions at time $t = 8\,P_{i}$.  
Top row: maps of (poloidal) plasma beta, $\beta \equiv p/(B_p^2/8\pi)$.  Poloidal magnetic field lines (magenta) and streamlines (white) are shown in relation to the critical velocity surfaces (dashed contours): the sonic surface (black), Alfv\'en surface (blue), and the fast and slow magnetosonic surfaces (red and cyan, respectively).  
Middle row: maps of $|B_\phi|/B_p$, with magnetic field lines and the Alfv\'en surface again overlaid along with contours of constant angular velocity, namely $\Omega(\varpi, \pi/2)$ evaluated at $\varpi = 3.0,3.5,4.0$ and $4.5$.    
Bottom row: forces along a streamline with footpoint at $(\varpi,z) = (3.5\,R_{IC}, 0.05\,R_{IC})$.  Dashed portions of lines indicate where the force is negative.  
Vertical dotted lines mark the crossing of the critical surfaces.}
\end{figure*}

\begin{figure*}
\includegraphics[width=0.9\textwidth]{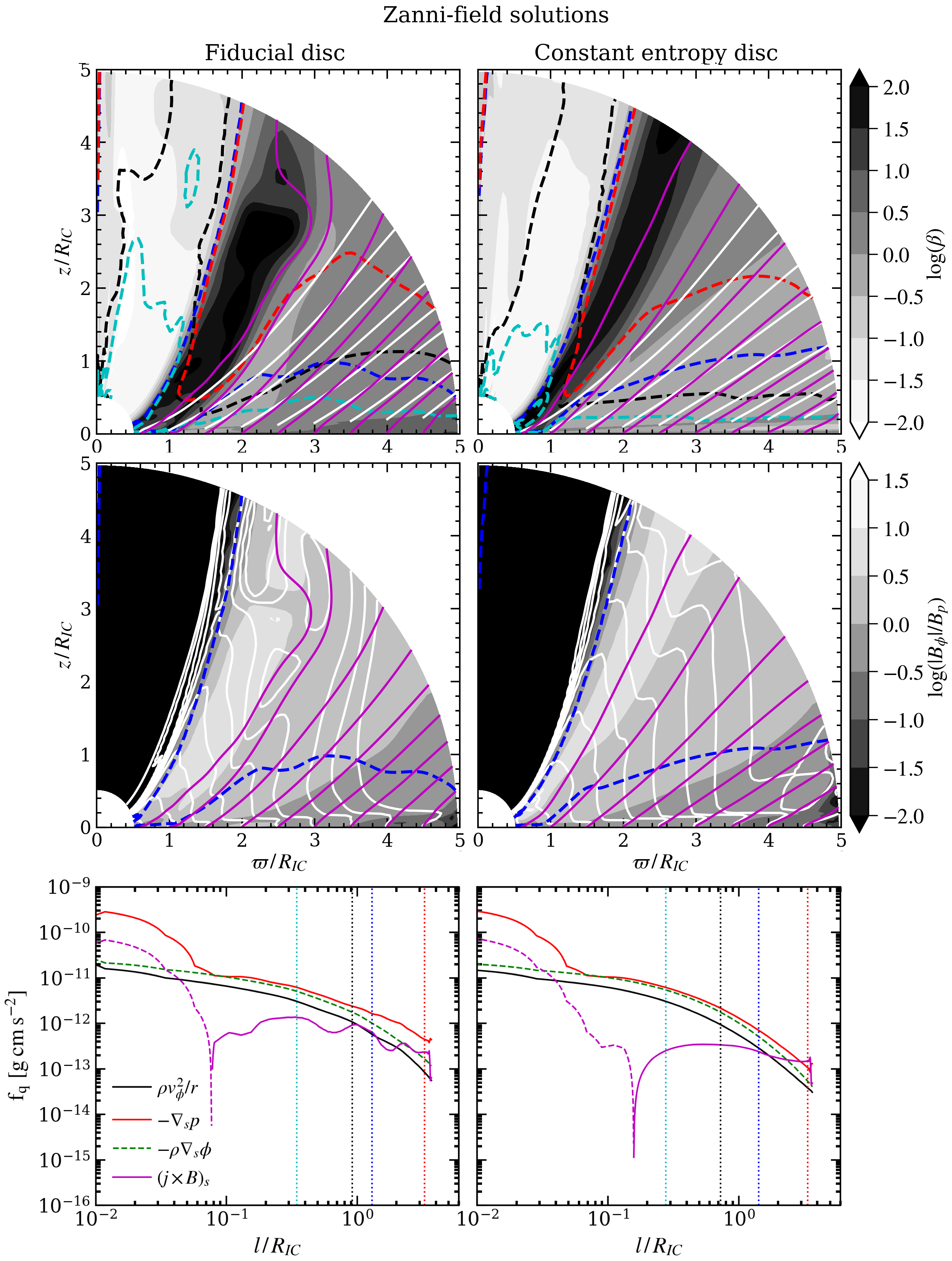}
\caption{
Same as Figure~7 but for the Zanni-field solutions.  Note that in the middle row, contours of specific angular velocity are now chosen by evalutating $\Omega$ along the midplane at every $0.5 R_{IC}$.  These nowhere align with the magnetic fields, indicating that magneto-centrifugal driving is absent.  In the bottom row, forces are shown along a streamline with footpoint at $(\varpi,z) = (1.5\,R_{IC}, 0.05\,R_{IC})$.  Notice the Lorentz force exceeds the gas pressure force at $\sim 3R_{IC}$ along the streamline for the constant entropy disk, which corresponds to the region showing enhanced wind velocity in Figure~5.}  
\end{figure*}

\begin{center}
\subsection{Magneto-centrifugal vs. magnetic pressure driving}
\end{center} 
The acceleration mechanism refers to forces able to accelerate the gas after it has been launched from the disk.  By design, the initial mass loading is done by the gas pressure force since a relatively strong thermally driven disk wind is always present.  As discussed in \S{2}, a magnetic field is capable of accelerating the flow either magneto-centrifugally or through the magnetic pressure gradient.  The latter force was defined as the projection of the poloidal component of the Lorentz force onto poloidal magnetic field lines, $F_\parallel$, while the former is due to the toroidal component of the Lorentz force, $F_\phi$.  
A map of the ratio $|B_\phi|/B_p = |F_\parallel|/F_\phi$ serves as a global diagnostic for assessing whether the magneto-centrifugal mechanism is operating: the toroidal force should dominate, implying a stronger poloidal field component.  Further diagnostics are suggested by the picture accompanying the Blandford-Payne model, namely that strong poloidal field lines are analogous to wires anchored in the disk and rigidly corotating with it up until about the Alfv\'en radius when the gas pressure dominates magnetic pressure.  This picture requires the poloidal plasma beta be much less than one and that contours of angular velocity be aligned with fieldlines within the Alfv\'en surface (e.g., Proga 2003).  

These diagnostics are shown in the top two rows of Figures~7 and 8 for the vertical field and Zanni-field runs, respectively.  
The top row of Figure~7 reveals that indeed $\beta << 1$ in the steady wind region, i.e. along streamlines emanating from $\varpi > 2 R_{IC}$ (for clarity, we only show the domain from $0$ to $\pi/2$).  We see from the middle panels that $B_\phi << B_p$ only for $\varpi \gtrsim 3.5 R_{IC}$.  Alignment of angular velocity contours with field lines is also seen for $\varpi \gtrsim 3.5 R_{IC}$ by overplotting contours of $\Omega = v_\phi/\varpi$ for values of $\Omega$ at $\varpi = 3$, 3.5, 4, and $4.5\,R_{IC}$ at the disk midplane.  We therefore find that this outermost wind region --- that showing significant velocity suppression --- has all the right conditions for magneto-centrifugal wind launching!  To better understand the gas dynamics of this wind we perform a detailed force analysis for the streamline with footpoint at $\varpi = 3.5 R_{IC}$.  This is shown in the bottom row of Figure~7.  Since poloidal magnetic field lines and streamlines are not perfectly aligned, we calculate the component of $\mathbf{j}\times \mathbf{B}/c$ along the streamline, denoted $(\mathbf{j}\times \mathbf{B})_s$, rather than $F_\parallel$.  These panels reveal that the dominant force by far is that due to gas pressure (dashed portions of lines indicate where a force is negative), with the centrifugal force balancing gravity within $0.1R_{IC}$ along the streamline but exceeding it at greater distances.  Notice that $(\mathbf{j}\times \mathbf{B})_s$ is mostly negligible owing to the small toroidal field in this region.  Thus, the Blandford-Payne mechanism appears to operate but is weak at such large radii due to relatively small rotational velocities.  The much stronger gas pressure gradient is still insufficient to provide the acceleration it did along the streamline at $\varpi = 3.5 R_{IC}$ in the pure thermal wind solution.  
We examine this issue further in \S{\ref{suppression}}.  
 
Recalling Figure~3, the Zanni-field is strong at small radii, while the vertical field is strong at large radii. 
We therefore do not expect the magneto-centrifugal mechanism to operate at large distances in the Zanni-field runs because $\beta$ is order unity in the steady wind region (see to top panels of Figure~8), while $B_\phi$ and $B_p$ are comparable according to the middle panels.  Magnetic pressure driving appears altogether absent in the vertical field solutions since $\beta$ is everywhere large in the unsteady, inner disk region where $B_\phi$ is large.  For the Zanni-field runs, the innermost streamlines within $2 R_{IC}$ occupy a region where both $B_\phi \gtrsim B_p$ and $\beta$ is order unity or larger, indicating that the increased velocity seen in Figure~5 could be due to a combination of magnetic and thermal driving.  
In the bottom panel of Figure~8 we examine the forces along a streamline with footpoint at $\varpi = 1.5 R_{IC}$, which is seen to cross all of the critical surfaces.  We again see that the gas pressure force is dominant at the base of the flow, but now $(\mathbf{j}\times \mathbf{B})_s$ becomes the largest force somewhat beyond the Alfv\'en radius in the case of the constant entropy disk, explaining why the peak velocities in Figure~5 are greater compared with those of our fiducial disk.  Thus, we have identified a clear circumstance of a magnetothermal effect leading to wind enhancement, the aiding force being the gradient of the toroidal magnetic pressure.   

\begin{figure*}
\includegraphics[width=0.7\textwidth]{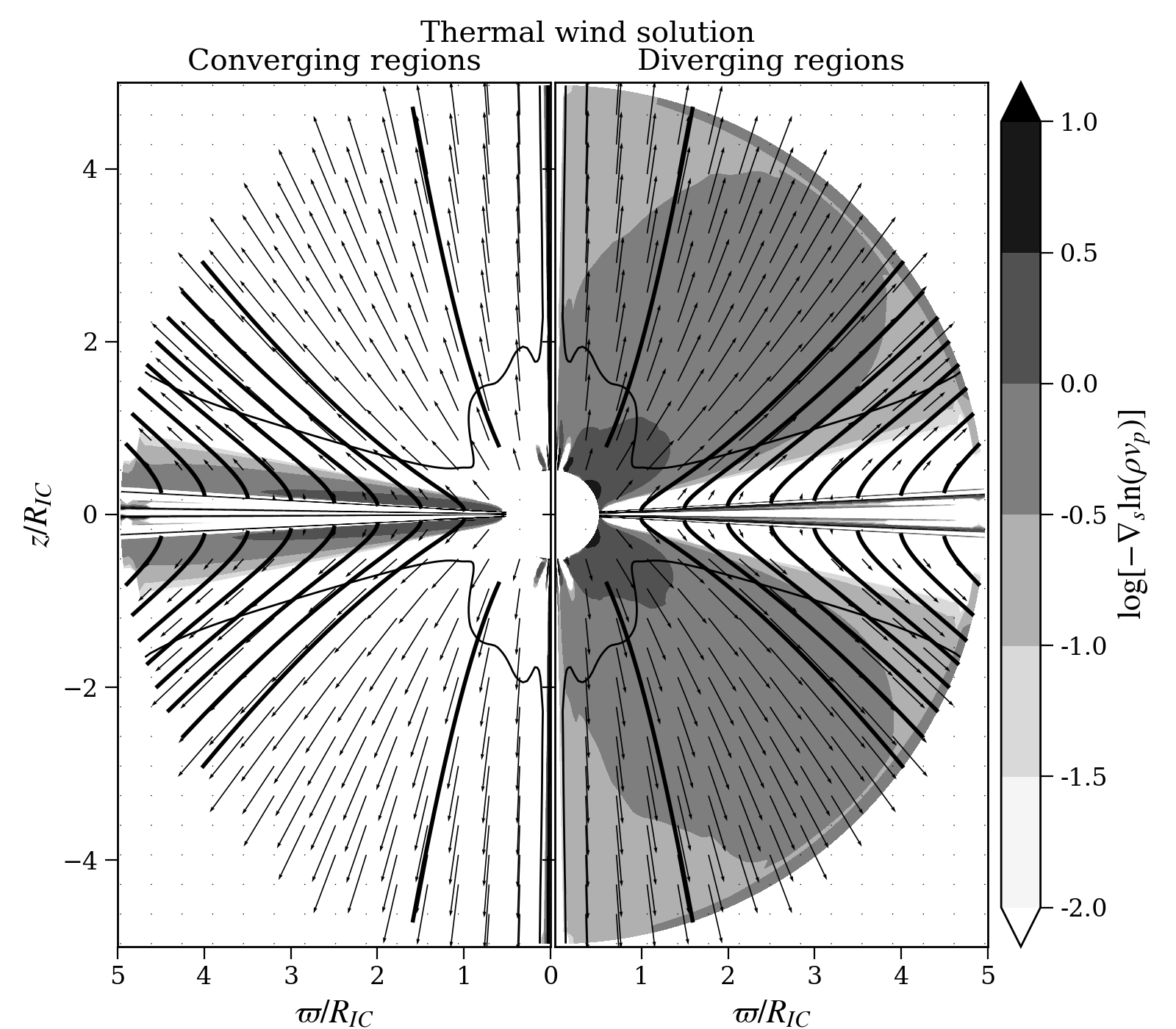}
\includegraphics[width=0.7\textwidth]{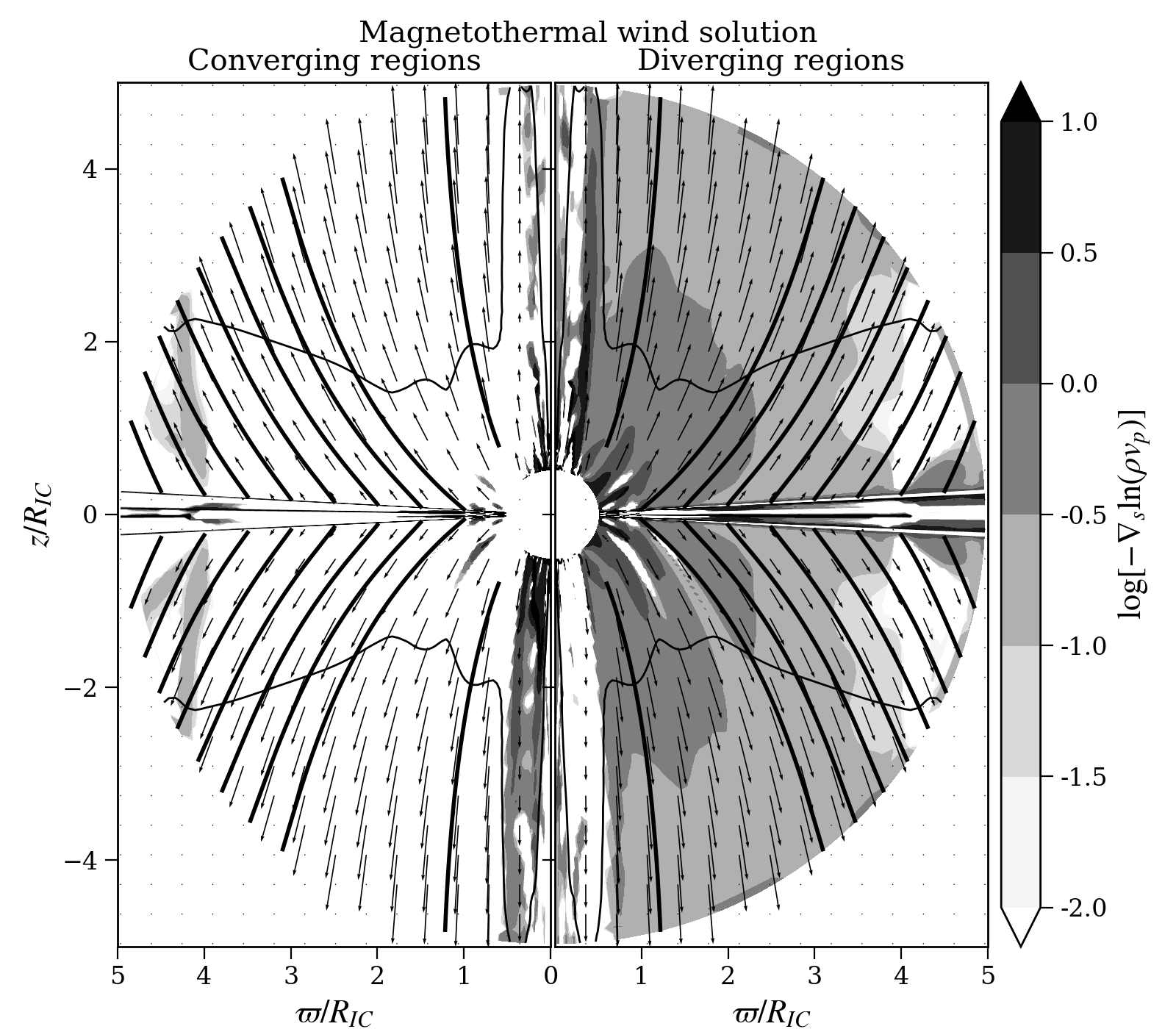}
\caption{
Colormaps of the quantitity $- \nabla_s \ln(\rho\,v_p)$, a diagnostic for assessing local flow convergence or divergence along streamlines, shown for the steady state thermal wind solution of our fiducial disk (top panels) and for the magnetothermal wind solution (at $t=4\,P_i$) of our fiducial disk with an initial vertical field.  Streamlines (black solid lines) are overplotted, as well as the sonic surface (thin solid line).  In a steady state, $- \nabla_s \ln(\rho\,v_p) = \nabla_s \ln(A)$.  Negative (positive) values of this quantity are plotted in the left (right) panels to indicate regions where the flow is converging (diverging).  The direct cause of the thermal wind suppression can be seen by comparing these maps: the acceleration region of the magnetothermal disk wind is no longer a converging-diverging geometry.}
\end{figure*}

\bigskip
\subsection{Disk wind suppression}
\label{suppression}
In addition to $F_\parallel$ and $F_\phi$, 
the streamlines are subject to a magnetic force acting perpendicular to magnetic surfaces in the poloidal plane, namely (Ferreira 1997), 
\begin{equation}
F_\perp = B_p j_\phi - \f{B_\phi}{2\pi r}\nabla_\perp I,
\end{equation}
where $\nabla_\perp$ is the gradient taken normal to the curve of the magnetic surface.  
This force is responsible for collimation, which will indirectly affect the acceleration of the flow by reshaping 
the `effective nozzle function' the flow traverses (e.g., Lamers \& Cassinelli 1999).  
This force is therefore indirectly responsible for the strong velocity suppression in the vertical field runs,
while it is weaker for the Zanni-field runs, and as a consequence the ram pressure of the thermal wind can overcome the tendency of the field to collimate.  
More simply, the poloidal plasma beta $\beta > 1$ in Figure~8 in the region $r > 2 R_{IC}$, while $\beta \sim 0.1$ for the vertical field runs as shown in Figure~7, implying that the magnetic field pressure is strong enough to control the flow geometry.  Indeed, animations of these plots reveal that the magnetic field always bends back after first being `blown' in the direction of the wind, but it bends back less so in the Zanni-field runs.  Once the magnetic field achieves cross-field balance and the streamlines have been reoriented to nearly align with it, $F_\perp$ is no longer a relevant force for understanding suppression of the thermal wind. 
That is, reorientation of the streamlines can make the flow susceptible to deceleration, but the actual cause is found by examining the net forces acting along the new streamline geometry --- and the geometry itself.   

We saw from Figure~7 that the dominant force is due to gas pressure, and its effectiveness depends on whether or not the flow traverses an effective converging-diverging geometry or something less conducive to acceleration.  To address this flow tube area argument more quantitatively, we examine a diagnostic derived from the steady state continuity equation, which can be written
\beq
\rho(l)\,v_p(l)\,A(l) = \dot{M},
\seq
where $l$ is the distance along the streamline, $v_p$ is the poloidal velocity along $l$, $A(l)$ is the flow area occupied along the streamline, and $\dot{M}$ is the local mass loss rate, a constant (for a derivation, see Waters \& Proga 2012).  Taking the gradient along $l$, denoted $\nabla_s$, gives 
\beq 
\nabla_s \ln(A) = - \nabla_s \ln(\rho\,v_p).
\label{dlnA}
\seq
This expression relates the changes in area occupied by the flow to the density and poloidal velocity.  The right hand side  can be computed globally, not just along a specific streamline, through the directional gradient $\nabla_s \equiv \hat{s} \cdot \nabla$, where $\hat{s}$ is the unit vector along the streamline.  For example, in spherical coordinates $\hat{s} = (v_r/v_p)\hat{r}  + (v_\theta/v_p)\hat{\theta}$, while $\nabla = (\partial/\partial r) \hat{r} + r^{-1}(\partial/\partial \theta) \hat{\theta}$ in axi-symmetry.  Equation \eqref{dlnA} is strictly valid only in a steady state, but the quantity $- \nabla_s \ln(\rho\,v_p)$ is a diagnostic of changes in the area along streamlines even in time-dependent flows.  

We calculate $- \nabla_s \ln(\rho\,v_p)$ for both the thermal and magnetothermal wind solutions using our fiducial disk in Figure~9.  This quantity can vary over several orders in magnitude and can change sign, so to plot it in log-scale we separately calculate its negative and positive values, which are shown on the left and right panels, respectively, for each solution.  Positive values indicate $dA > 0$, i.e. the flow is diverging along the streamline.  We see that the thermal wind solution transitions from a converging to diverging geometry interior to the sonic surface.  (In the absence of gravity, the transition should happen at the sonic surface, while here it is seen to be somewhat interior.)  We know in the case of a Parker wind that transonic flows in the presence of gravity can be purely diverging.  Indeed, the radial portions of the thermal wind (the flow within the second streamline in the top panels of Figure~9) are always diverging.  The geometry of the magnetothermal wind, on the other hand, is quite different, as the flow is no longer converging as it rises above the disk surface.  Despite there still being strong pressure gradients along these streamlines, with magnitudes that can even exceed  those along the thermal wind streamlines, independent geometric considerations indicate that the conditions are not as ideal for accelerating the outflow.  

\section{Discussion and conclusions}
Magnetically launched winds require relatively strong magnetic fields ($\beta \lesssim 1$), yet we have shown in this paper that such field strengths can actually suppress the thermal driving mechanism.  The reason is that when field lines are strong enough to control the directionality of the flow, the streamline geometry imposes geometric constraints that may prevent the gas pressure force from being optimally effective at providing acceleration.  
In summary, we have identified two magnetothermal effects: (i) wind suppression arising from a reorientation of the flow streamlines in low$-\beta$ regions with $B_\phi << B_p$ and (ii) wind enhancement due to magnetic pressure driving in regions with $\beta \sim 1$ and $B_\phi \sim B_p$.  The second effect only occurs over a narrow solid angle in our Zanni-field runs and thus is insignificant for the overall kinetic luminosity of the wind, while the first effect occurs over a large solid angle in our vertical field runs, thereby reducing the kinetic luminosity substantially.  We do not expect these results to change in resistive MHD treatments of this problem, as resistivity is typically only important inside the disk.

Since our simulations were designed to explore the optimal radii for thermal wind launching, we conclude that magnetothermal winds are less powerful than pure thermal winds near the Compton radius, at least for the field strengths we were able to explore.  Our findings are suggestive of a very compact origin for the outflow observed in GRO J1655-40, well within the Compton radius where thermal driving is strongly inhibited by gravity but where magnetic flux can accumulate and lead to more highly magnetized disks.  This conclusion is supported by the findings of Stepanovs \& Fendt (2016), who showed (using the constant entropy disk, albeit with resistivity) that higher disk magnetization leads to faster outflows.  It is also supported by the recent \emph{Chandra} observations of GRS 1915+105 reported by Miller et al. (2016), who by assuming equipartition between the magnetic field pressure and either the gas pressure (in the case of magnetic pressure driving) or the wind ram pressure (in the case of magneto-centrifugal driving) inferred magnetic field strengths of at least $10^3\,\rm{G}$ or $10^4\,\rm{G}$, respectively.  The basic requirement that the rotational energy of the disk exceed the magnetic energy places the following upper bound on the wind location,
\beq
r_{\rm{wind}} <  2\pi \mu m_p\,c^2 \f{n}{B^2} R_S = 9.5\times10^3 \f{n_{15}}{B_4^2} R_S,
\seq
where $n_{15}$ is the number density in units of $10^{15}\,\rm{cm}^{-3}$ and $B_4$ is the field strength in units of $10^4\,\rm{G}$, as appropriate for the wind in GRS 1915+105.
Comparison with equation \eqref{R_IC} shows this limit to be within $0.03\,R_{IC}$, assuming our fiducial Compton temperature $T_{IC} = 1.4\times10^7\rm{K}$.   

Independent considerations demand that a viable magnetothermal wind model tailored to GRO J1655-40 have an outer disk boundary at or within our current \emph{inner boundary} at $\sim 0.5 R_{IC}$.  
For a binary separation distance $D$ and mass ratio $q = M_{\rm{sec}}/M_{BH}$, Higginbottom et al. (2017) estimate the maximum allowable disk size as 
\beq
R_{d,\rm{max}}= \f{0.6}{1 + q}D,
\label{Rmax}
\seq
which is the radius where a test particle orbit is no longer intersecting (Warner 2003).  For a black hole mass of $7M_\odot$, $q = 0.4$, and $D = 7.5\times10^{11} \rm{cm}$ (roughly the parameters for GRO J1655-40), $R_{d,\rm{max}} \approx 3.2\times10^{11} \rm{cm}$.  From equation \eqref{R_IC} with $T_{IC} = 1.4\times10^7\rm{K}$, the Compton radius actually lies somewhat beyond $R_{d,\rm{max}}$ at $R_{IC} =  4.8\times10^{11} \rm{cm}$.  A realistic model for GRO J1655-40 should therefore use a disk with an outer boundary no greater than about $0.7R_{IC}$.  Our test runs have shown this to be a much more challenging region to explore numerically, the primary reason being that material launched from regions deeper in the gravitational potential well have a greater tendency to fallback toward the disk, thereby twisting the magnetic fields and generating turbulence.  Insights gained from steady state ideal MHD theory will be of less use for understanding such solutions.   
Moreover, under these circumstances, a simplified disk setup such as the one we use here would, given sufficient resolution, inevitably evolve to become a fully turbulent MRI disk.  Then the disk no longer serves as simply a reservoir of gas anchoring magnetic field lines, so the accretion disk physics must be carefully treated along with the wind physics.  
Thankfully, this theoretical effort should be greatly aided by further constraints provided by observations from \emph{Chandra} and \emph{XXM-Newton}, as well as new facilities such as \emph{NICER} (e.g., Neilsen et al. 2018).  

While 3D isothermal simulations of winds from MRI disks are becoming feasible (e.g., Ju, Stone \& Zhu 2017; Zhu \& Stone 2018), as well as 3D versions of the constant entropy disk setup including resistivity and accounting for the motion of the companion (Sheikhnezami \& Fendt 2018), it should be emphasized that the present 2D axisymmetric magnetothermal wind simulations are the first of their kind applied to LMXBs.  As such, it is important to fully understand the interplay of magnetic effects with various realistic heating and cooling functions tailored to X-ray binaries such as those recently considered in the 1D simulations of Dyda et al. (2017).  In particular, it is important to better understand the so-called over-ionization problem (e.g., Murray et al. 1994; Higginbottom et al. 2014) that leads to the common assumption that the winds in LMXBs cannot be line-driven due to a lack of strongly resonant ions (e.g., Proga \& Kallman 2002).  The recent work of Dannen et al. (in prep.) using \textsc{XSTAR} reveals that iron opacities can be very significant at the high ionization parameters characterizing LMXBs, implying that the line-driving mechanism may be viable after all.  Intriguingly, since the outer regions of X-ray binary disks are cool enough that the iron opacity bump (relevant in massive star envelopes [e.g., Paxton et al. 2013 and references therein] and AGN disks [Jiang et al. 2016]) may be important for setting the disk's vertical structure, the subsequent depletion of the iron abundance due a disk wind may subject the disk to thermal runaway if it is radiation-pressure dominated (e.g., Jiang et al. 2013, 2016, although see also Grz{\c e}dzielski et al. 2017).  Therefore, simulations of this combined disk-plus-wind physics may lead to models that naturally explain state transitions back to the low-hard state (see e.g., Fender et al. 2004; D{\'{\i}}az Trigo et al. 2014).   

\section{Acknowledgements}
This  work  was  supported  by  NASA  under  ATP  grant
NNX14AK44G.  We thank Jonathan Ferreira and Christian Fendt for stimulating discussions and email correspondence, as well as
Zhaohuan Zhu for frequent discussions and guidance with \athena during the early phases of this project.  
TW acknowledges a insightful meeting with Petros Tzeferacos during his visit to LANL
and useful discussions with Sergei Dyda, Hui Li, Yaping Li, Edison Liang, and Peter Polko on various aspects of MHD wind theory.  
 
\appendix
\section{\\Boundary conditions \& floors}
Our simulations extend all the way to the poles at $\theta=0$ and $\pi$, made possible by a special polar boundary condition that was implemented into \athena to prevent the loss of magnetic
fields that can occur when the poles are excluded from the domain (see Zhu \& Stone 2018).  In 2D axisymmetric setups, this is almost identical to a reflecting boundary condition and referred to in the input file as $\mathtt{polar\_wedge}$.  The midplane BCs discussed in \S{2.3} are enforced as an `immersed' boundary condition since they apply to active zones instead of ghost zones (GZs).  Specifically, the conserved variables in the first active zone above and below $90^\circ$ (here defined $\mathtt{j_{\rm{mid}}}$ and $\mathtt{j_{\rm{mid}+1}}$) are reset to the midplane profiles defined in equations (3) or (6).  
This is done via a user enrolled boundary function inside the problem generator, which is called after both a half and a full timestep.  
No such midplane BC is applied to the magnetic field: the conserved energy, $p/(\gamma - 1) + \rho v^2/2 + B^2/8\pi$, is first reset to $p/(\gamma - 1) + \rho v_\phi^2/2$ at $\mathtt{j_{\rm{mid}}}$ and $\mathtt{j_{\rm{mid}+1}}$ according to equations (3) or (6) and then $B^2/8\pi$ is added to this quantity.  Note that a factor of $\sqrt{4\pi}$ is absorbed into the magnetic field in \athenas code units, i.e. the conserved energy is $p/(\gamma - 1) + \rho v^2/2 + B^2/2$ using \athenas $B$.  

For the outer radial boundary, we apply custom boundary conditions designed to minimize the numerical effect of artificial collimation discussed in detail by Ustyugova et al. (1999) and Zanni et al. (2007).   This tendency of poloidal field lines touching the outer boundary to vertically collimate results from unbalanced Lorentz forces when zero-gradient boundary conditions (i.e. standard outflow BCs) are employed on all variables.   
In our test runs attempting standard outflow BCs, this effect is observed after just a few inner orbital times.  Our custom BCs 
instead apply \emph{constant-gradient} BCs to $B_\theta$ and $B_\phi$: we linearly interpolate the two last active zones into the outer GZs.  As discussed by Zanni et al. (2007), this constant-gradient BC must also be applied to $v_\phi$, while the solenoidal condition determines $B_r$.  In \athena, enforcing $\nabla \cdot \mathbf{B} = 0$ is internally handled by the CT-scheme, so all that needs to be done for $B_r$ in the custom BC is to set its value in the second GZ to that of the first GZ, leaving the first GZ unspecified.  Standard outflow BCs are applied to all other variables $(\rho, v_r, v_\theta, p)$, except $v_r$ is set to zero in the GZs if $v_r < 0$ in the last active zone; this prevents a Bondi-type inflow from developing.  

More freedom is allowed at the inner radial boundary, but an analogous BC for $r_{\rm{in}}$ was found to be very robust, except we additionally apply constant-gradient BCs to both density and pressure and no longer test if $v_r < 0$.    
These custom BCs allow us to evolve the solutions to about $10 P_i$, beyond which time only the constant entropy disk/Zanni-field solution continues to remain unsusceptible to artificial collimation.  The behavior of the other three runs beyond $t=8\,P_i$ in Figure~6 does not appear to be a consequence of the artificial collimation but rather the trigger: episodic ejections of disk material that then outflow through the outer boundary have adverse effects on maintaining the solutions.  

The default floor routine in \athena was also customized.  The pressure floor was set to $p_{\rm{min}} = 10^{-5}\rho_0 v_{\rm{kep}}^2(R_{IC})$, but instead of a constant density floor, we use a profile, $\rho_{\rm{min}}(r) = 2.5\times10^{-9} (r_{\rm{in}}/r)^2$.  We additionally track the triggering of floors in either $\rho$ or $p$.  The source term $\rho\mathcal{L}$ in equation (13) is solved semi-implicitly as described in the appendix of Dyda et al. (2017), and this involves a root solve that can fail if the zone being updated triggered a floor in the hydro or MHD modules.  We therefore added a 2D array of bools to the mesh structure in \athena and only attempted the semi-implicit update if no floors were triggered, using a fully explicit update otherwise.


\begin{thebibliography}{}
\bibitem[Anderson et al.(2005)]{2005ApJ...630..945A} Anderson, J.~M., Li, Z.-Y., Krasnopolsky, R., \& Blandford, R.~D.\ 2005, \apj, 630, 945
\bibitem[Bai et al.(2016)]{2016ApJ...818..152B} Bai, X.-N., Ye, J., Goodman, J., \& Yuan, F.\ 2016, \apj, 818, 152
\bibitem[Begelman et al.(1983)]{1983ApJ...271...70B} Begelman, M.~C., McKee, C.~F., \& Shields, G.~A.\ 1983, \apj, 271, 70
\bibitem[Belcher \& MacGregor(1976)]{1976ApJ...210..498B} Belcher, J.~W., \& MacGregor, K.~B.\ 1976, \apj, 210, 498
\bibitem[Belloni(2010)]{2010LNP...794...53B} Belloni, T.~M.\ 2010, Lecture Notes in Physics, Berlin Springer Verlag, 794, 53
\bibitem[Blandford \& Payne(1982)]{1982MNRAS.199..883B} Blandford, R.~D., \& Payne, D.~G.\ 1982, \mnras, 199, 883
\bibitem[Blondin(1994)]{1994ApJ...435..756B} Blondin, J.~M.\ 1994, \apj, 435, 756
\bibitem[Bogovalov(1997)]{1997A&A...323..634B} Bogovalov, S.~V.\ 1997, \aap, 323, 634 
\bibitem[Cassinelli(1990)]{1990ASIC..316..135C} Cassinelli, J.\ 1990, NATO Advanced Science Institutes (ASI) Series C, 316, 135 
\bibitem[Chakravorty et al.(2016)]{2016A&A...589A.119C} Chakravorty, S., Petrucci, P.-O., Ferreira, J., et al.\ 2016, \aap, 589, A119
\bibitem[Clarke \& Alexander(2016)]{2016MNRAS.460.3044C} Clarke, C.~J., \& Alexander, R.~D.\ 2016, \mnras, 460, 3044
\bibitem[Contopoulos(1996)]{1996ApJ...460..185C} Contopoulos, J.\ 1996, \apj, 460, 185 
\bibitem[D{\'{\i}}az Trigo et al.(2007)]{2007A&A...462..657D} D{\'{\i}}az Trigo, M., Parmar, A.~N., Miller, J., Kuulkers, E., \& Caballero-Garc{\'{\i}}a, M.~D.\ 2007, \aap, 462, 657
\bibitem[D{\'{\i}}az Trigo \& Boirin(2013)]{2013AcPol..53..659D} D{\'{\i}}az Trigo, M., \& Boirin, L.\ 2013, Acta Polytechnica, 53, 659
\bibitem[D{\'{\i}}az Trigo et al.(2014)]{2014A&A...571A..76D} D{\'{\i}}az Trigo, M., Migliari, S., Miller-Jones, J.~C.~A., \& Guainazzi, M.\ 2014, \aap, 571, A76
\bibitem[Done et al.(2007)]{2007A&ARv..15....1D} Done, C., Gierli{\'n}ski, M., \& Kubota, A.\ 2007, \aapr, 15, 1
\bibitem[Dorodnitsyn et al.(2008)]{2008ApJ...687...97D} Dorodnitsyn, A., Kallman, T., \& Proga, D.\ 2008, \apj, 687, 97
\bibitem[Dyda et al.(2017)]{2017MNRAS.467.4161D} Dyda, S., Dannen, R., Waters, T., \& Proga, D.\ 2017, \mnras, 467, 4161
\bibitem[Fender et al.(2004)]{2004MNRAS.355.1105F} Fender, R.~P., Belloni, T.~M., \& Gallo, E.\ 2004, \mnras, 355, 110
\bibitem[Fender \& Gallo(2014)]{2014SSRv..183..323F} Fender, R., \& Gallo, E.\ 2014, \ssr, 183, 323
\bibitem[Fendt \& Sheikhnezami(2013)]{2013ApJ...774...12F} Fendt, C., \& Sheikhnezami, S.\ 2013, \apj, 774, 12
\bibitem[Fendt \& Ga{\ss}mann(2018)]{2018ApJ...855..130F} Fendt, C., \& Ga{\ss}mann, D.\ 2018, \apj, 855, 130
\bibitem[Ferreira(1997)]{1997A&A...319..340F} Ferreira, J.\ 1997, \aap, 319, 340
\bibitem[Ferreira et al.(2006)]{2006A&A...447..813F} Ferreira, J., Petrucci, P.-O., Henri, G., Saug{\'e}, L., \& Pelletier, G.\ 2006, \aap, 447, 813 
\bibitem[Ferreira(2007)]{2007LNP...723..181F} Ferreira, J.\ 2007, Lecture Notes in Physics, Berlin Springer Verlag, 723, 181
\bibitem[Font et al.(2004)]{2004ApJ...607..890F} Font, A.~S., McCarthy, I.~G., Johnstone, D., \& Ballantyne, D.~R.\ 2004, \apj, 607, 890
\bibitem[Fukumura et al.(2017)]{2017NatAs...1E..62F} Fukumura, K., Kazanas, D., Shrader, C., et al.\ 2017, Nature Astronomy, 1, 0062
\bibitem[Gorti et al.(2016)]{2016SSRv..205..125G} Gorti, U., Liseau, R., S{\'a}ndor, Z., \& Clarke, C.\ 2016, \ssr, 205, 125
\bibitem[Grz{\c e}dzielski et al.(2017)]{2017ApJ...845...20G} Grz{\c e}dzielski, M., Janiuk, A., \& Czerny, B.\ 2017, \apj, 845, 20
\bibitem[Hawley et al.(2015)]{2015SSRv..191..441H} Hawley, J.~F., Fendt, C., Hardcastle, M., Nokhrina, E., \& Tchekhovskoy, A.\ 2015, \ssr, 191, 441 
\bibitem[Higginbottom et al.(2014)]{2014ApJ...789...19H} Higginbottom, N., Proga, D., Knigge, C., et al.\ 2014, \apj, 789, 19
\bibitem[Higginbottom \& Proga(2015)]{2015ApJ...807..107H} Higginbottom, N., \& Proga, D.\ 2015, \apj, 807, 107
\bibitem[Higginbottom et al.(2017)]{2017ApJ...836...42H} Higginbottom, N., Proga, D., Knigge, C., \& Long, K.~S.\ 2017, \apj, 836, 42 
\bibitem[Higginbottom et al.(2018)]{2018arXiv180604887H} Higginbottom, N., Knigge, C., Long, K.~S., et al.\ 2018, arXiv:1806.04887
\bibitem[Janiuk et al.(2015)]{2015A&A...574A..92J} Janiuk, A., Grzedzielski, M., Capitanio, F., \& Bianchi, S.\ 2015, \aap, 574, A92
\bibitem[Jiang et al.(2013)]{2013ApJ...778...65J} Jiang, Y.-F., Stone, J.~M., \& Davis, S.~W.\ 2013, \apj, 778, 6574
\bibitem[Jiang et al.(2016)]{2016ApJ...827...10J} Jiang, Y.-F., Davis, S.~W., \& Stone, J.~M.\ 2016, \apj, 827, 10 
\bibitem[Jiang et al.(2017)]{2017arXiv170902845J} Jiang, Y.-F., Stone, J., \& Davis, S.~W.\ 2017, arXiv:1709.02845
\bibitem[Ju et al.(2017)]{2017ApJ...841...29J} Ju, W., Stone, J.~M., \& Zhu, Z.\ 2017, \apj, 841, 29 
\bibitem[Kallman et al.(2009)]{2009ApJ...701..865K} Kallman, T.~R., Bautista, M.~A., Goriely, S., et al.\ 2009, \apj, 701, 865
\bibitem[King et al.(2012)]{2012ApJ...746L..20K} King, A.~L., Miller, J.~M., Raymond, J., et al.\ 2012, \apjl, 746, L20
\bibitem[King et al.(2013)]{2013ApJ...762..103K} King, A.~L., Miller, J.~M., Raymond, J., et al.\ 2013, \apj, 762, 103
\bibitem[Konigl \& Pudritz(2000)]{2000prpl.conf..759K} Konigl, A., \& Pudritz, R.~E.\ 2000, Protostars and Planets IV, 759
\bibitem[K{\"o}nigl \& Salmeron(2011)]{2011ppcd.book..283K} K{\"o}nigl, A., \& Salmeron, R.\ 2011, Physical Processes in Circumstellar Disks around Young Stars, 283
\bibitem[Kubota et al.(2007)]{2007PASJ...59S.185K} Kubota, A., Dotani, T., Cottam, J., et al.\ 2007, \pasj, 59, 185
\bibitem[Kurosawa \& Proga(2009)]{2009ApJ...693.1929K} Kurosawa, R., \& Proga, D.\ 2009, \apj, 693, 1929 
\bibitem[Lamers \& Cassinelli(1999)]{1999isw..book.....L} Lamers, H.~J.~G.~L.~M., \& Cassinelli, J.~P.\ 1999, Introduction to Stellar Winds, by Henny J.~G.~L.~M.~Lamers and Joseph P.~Cassinelli, pp.~452.~ISBN 0521593980.~Cambridge, UK: Cambridge University Press, June 1999., 452
\bibitem[Lee et al.(2002)]{2002ApJ...567.1102L} Lee, J.~C., Reynolds, C.~S., Remillard, R., et al.\ 2002, \apj, 567, 1102 
\bibitem[Livio(1997)]{1997ASPC..121..845L} Livio, M.\ 1997, IAU Colloq.~163: Accretion Phenomena and Related Outflows, 121, 845
\bibitem[Lovelace et al.(1986)]{1986ApJS...62....1L} Lovelace, R.~V.~E., Mehanian, C., Mobarry, C.~M., \& Sulkanen, M.~E.\ 1986, \apjs, 62, 1
\bibitem[Luketic et al.(2010)]{2010ApJ...719..515L} Luketic, S., Proga, D., Kallman, T.~R., Raymond, J.~C., \& Miller, J.~M.\ 2010, \apj, 719, 515
\bibitem[Marcel et al.(2018)]{2018arXiv180512407M} Marcel, G., Ferreira, J., Petrucci, P., et al.\ 2018, arXiv:1805.12407
\bibitem[Michel(1969)]{1969ApJ...158..727M} Michel, F.~C.\ 1969, \apj, 158, 727
\bibitem[Miller et al.(2006)]{2006ApJ...646..394M} Miller, J.~M., Raymond, J., Homan, J., et al.\ 2006, \apj, 646, 394
\bibitem[Miller et al.(2006)]{2006Natur.441..953M} Miller, J.~M., Raymond, J., Fabian, A., et al.\ 2006, \nat, 441, 953
\bibitem[Miller et al.(2008)]{2008ApJ...680.1359M} Miller, J.~M., Raymond, J., Reynolds, C.~S., et al.\ 2008, \apj, 680, 1359
\bibitem[Miller et al.(2014)]{2014ApJ...788...53M} Miller, J.~M., Raymond, J., Kallman, T.~R., et al.\ 2014, \apj, 788, 53
\bibitem[Miller et al.(2016)]{2016ApJ...821L...9M} Miller, J.~M., Raymond, J., Fabian, A.~C., et al.\ 2016, \apjl, 821, L9
\bibitem[Mo{\'s}cibrodzka \& Proga(2013)]{2013ApJ...767..156M} Mo{\'s}cibrodzka, M., \& Proga, D.\ 2013, \apj, 767, 156
\bibitem[Murphy et al.(2010)]{2010A&A...512A..82M} Murphy, G.~C., Ferreira, J., \& Zanni, C.\ 2010, \aap, 512, A82
\bibitem[Murray et al.(1994)]{1994ApJ...435..631M} Murray, S.~D., Castor, J.~I., Klein, R.~I., \& McKee, C.~F.\ 1994, \apj, 435, 631
\bibitem[Neilsen \& Lee(2009)]{2009Natur.458..481N} Neilsen, J., \& Lee, J.~C.\ 2009, \nat, 458, 481
\bibitem[Neilsen \& Homan(2012)]{2012ApJ...750...27N} Neilsen, J., \& Homan, J.\ 2012, \apj, 750, 27
\bibitem[Neilsen(2013)]{2013AdSpR..52..732N} Neilsen, J.\ 2013, Advances in Space Research, 52, 732
\bibitem[Neilsen et al.(2016)]{2016ApJ...822...20N} Neilsen, J., Rahoui, F., Homan, J., \& Buxton, M.\ 2016, \apj, 822, 20
\bibitem[Neilsen et al.(2018)]{2018arXiv180602342N} Neilsen, J., Cackett, E., Remillard, R.~A., et al.\ 2018, arXiv:1806.02342
\bibitem[Nerney(1980)]{1980ApJ...242..723N} Nerney, S.\ 1980, \apj, 242, 723
\bibitem[Netzer(2006)]{2006ApJ...652L.117N} Netzer, H.\ 2006, \apjl, 652, L117
\bibitem[Ouyed \& Pudritz(1997)]{1997ApJ...484..794O} Ouyed, R., \& Pudritz, R.~E.\ 1997, \apj, 484, 794 
\bibitem[Ouyed \& Pudritz(1999)]{1999MNRAS.309..233O} Ouyed, R., \& Pudritz, R.~E.\ 1999, \mnras, 309, 233
\bibitem[Paxton et al.(2013)]{2013ApJS..208....4P} Paxton, B., Cantiello, M., Arras, P., et al.\ 2013, \apjs, 208, 4
\bibitem[Poe et al.(1989)]{1989ApJ...337..888P} Poe, C.~H., Friend, D.~B., \& Cassinelli, J.~P.\ 1989, \apj, 337, 888 
\bibitem[Ponti et al.(2012)]{2012MNRAS.422L..11P} Ponti, G., Fender, R.~P., Begelman, M.~C., et al.\ 2012, \mnras, 422, L11
\bibitem[Proga et al.(2000)]{2000ApJ...543..686P} Proga, D., Stone, J.~M., \& Kallman, T.~R.\ 2000, \apj, 543, 686
\bibitem[Proga \& Kallman(2002)]{2002ApJ...565..455P} Proga, D., \& Kallman, T.~R.\ 2002, \apj, 565, 455
\bibitem[Proga(2003)]{2003ApJ...585..406P} Proga, D.\ 2003, \apj, 585, 406
\bibitem[Proga \& Waters(2015)]{2015ApJ...804..137P} Proga, D., \& Waters, T.\ 2015, \apj, 804, 137
\bibitem[Pudritz \& Norman(1986)]{1986ApJ...301..571P} Pudritz, R.~E., \& Norman, C.~A.\ 1986, \apj, 301, 571
\bibitem[Pudritz et al.(2007)]{2007prpl.conf..277P} Pudritz, R.~E., Ouyed, R., Fendt, C., \& Brandenburg, A.\ 2007, Protostars and Planets V, 277
\bibitem[Sauty et al.(2002)]{2002LNP...589...41S} Sauty, C., Tsinganos, K., \& Trussoni, E.\ 2002, Relativistic Flows in Astrophysics, 589, 41
\bibitem[Sheikhnezami et al.(2012)]{2012ApJ...757...65S} Sheikhnezami, S., Fendt, C., Porth, O., Vaidya, B., \& Ghanbari, J.\ 2012, \apj, 757, 65
\bibitem[Sheikhnezami \& Fendt(2015)]{2015ApJ...814..113S} Sheikhnezami, S., \& Fendt, C.\ 2015, \apj, 814, 113
\bibitem[Sheikhnezami \& Fendt(2018)]{2018arXiv180505962S} Sheikhnezami, S., \& Fendt, C.\ 2018, arXiv:1805.05962
\bibitem[Schulz \& Brandt(2002)]{2002ApJ...572..971S} Schulz, N.~S., \& Brandt, W.~N.\ 2002, \apj, 572, 971
\bibitem[Shidatsu et al.(2016)]{2016ApJ...823..159S} Shidatsu, M., Done, C., \& Ueda, Y.\ 2016, \apj, 823, 159
\bibitem[Spruit(1996)]{1996ASIC..477..249S} Spruit, H.~C.\ 1996, NATO Advanced Science Institutes (ASI) Series C, 477, 249
\bibitem[Stepanovs \& Fendt(2014)]{2014ApJ...793...31S} Stepanovs, D., \& Fendt, C.\ 2014, \apj, 793, 31
\bibitem[Stepanovs et al.(2014)]{2014ApJ...796...29S} Stepanovs, D., Fendt, C., \& Sheikhnezami, S.\ 2014, \apj, 796, 29
\bibitem[Stepanovs \& Fendt(2016)]{2016ApJ...825...14S} Stepanovs, D., \& Fendt, C.\ 2016, \apj, 825, 14
\bibitem[Stone et al.(2008)]{2008ApJS..178..137S} Stone, J.~M., Gardiner, T.~A., Teuben, P., Hawley, J.~F., \& Simon, J.~B.\ 2008, \apjs, 178, 137
\bibitem[Townsend(2009)]{2009ApJS..181..391T} Townsend, R.~H.~D.\ 2009, \apjs, 181, 391
\bibitem[Tsinganos(2007)]{2007LNP...723..117T} Tsinganos, K.\ 2007, Lecture Notes in Physics, Berlin Springer Verlag, 723, 117
\bibitem[Tzeferacos et al.(2009)]{2009MNRAS.400..820T} Tzeferacos, P., Ferrari, A., Mignone, A., et al.\ 2009, \mnras, 400, 820
\bibitem[Tzeferacos et al.(2013)]{2013MNRAS.428.3151T} Tzeferacos, P., Ferrari, A., Mignone, A., et al.\ 2013, \mnras, 428, 3151
\bibitem[Uchida \& Shibata(1985)]{1985PASJ...37..515U} Uchida, Y., \& Shibata, K.\ 1985, \pasj, 37, 515
\bibitem[Ueda et al.(2009)]{2009ApJ...695..888U} Ueda, Y., Yamaoka, K., \& Remillard, R.\ 2009, \apj, 695, 888
\bibitem[Ustyugova et al.(1999)]{1999ApJ...516..221U} Ustyugova, G.~V., Koldoba, A.~V., Romanova, M.~M., Chechetkin, V.~M., \& Lovelace, R.~V.~E.\ 1999, \apj, 516, 221
\bibitem[Uttley \& Klein-Wolt(2015)]{2015MNRAS.451..475U} Uttley, P., \& Klein-Wolt, M.\ 2015, \mnras, 451, 475
\bibitem[Warner(2003)]{2003cvs..book.....W} Warner, B.\ 2003, Cataclysmic Variable Stars, by Brian Warner, pp.~592.~ISBN 052154209X.~Cambridge, UK: Cambridge University Press, September 2003., 592
\bibitem[Waters \& Proga(2012)]{2012MNRAS.426.2239W} Waters, T.~R., \& Proga, D.\ 2012, \mnras, 426, 2239
\bibitem[Weber \& Davis(1967)]{1967ApJ...148..217W} Weber, E.~J., \& Davis, L., Jr.\ 1967, \apj, 148, 217
\bibitem[Woods et al.(1996)]{1996ApJ...461..767W} Woods, D.~T., Klein, R.~I., Castor, J.~I., McKee, C.~F., \& Bell, J.~B.\ 1996, \apj, 461, 767 
\bibitem[Zanni et al.(2007)]{2007A&A...469..811Z} Zanni, C., Ferrari, A., Rosner, R., Bodo, G., \& Massaglia, S.\ 2007, \aap, 469, 811
\bibitem[Zanni \& Ferreira(2009)]{2009A&A...508.1117Z} Zanni, C., \& Ferreira, J.\ 2009, \aap, 508, 1117
\bibitem[Zhang(2013)]{2013FrPhy...8..630Z} Zhang, S.-N.\ 2013, Frontiers of Physics, 8, 630
\bibitem[Zhu \& Stone(2018)]{2018ApJ...857...34Z} Zhu, Z., \& Stone, J.~M.\ 2018, \apj, 857, 34 
\end{thebibliography}
\end{document}